# Revealing the nontrivial topological surface states of catalysts for effective photochemical carbon dioxide conversion

*Kangwang Wang*[a], *Longfu Li*[a], *Peifeng Yu*[a], *Nannan Tang*[b], *Lingyong Zeng*[a], *Kuan Li*[a], *Chao Zhang*[a], *Rui Chen*[a], *Zaichen Xiang*[a], *Huichao Wang*[b], *Yongqing Cai*[c,*], *Kai Yan*[d,*], *Huixia Luo*[a,*]

[a] School of Materials Science and Engineering, State Key Laboratory of Optoelectronic Materials and Technologies, Key Lab of Polymer Composite & Functional Materials, Guangzhou Key Laboratory of Flexible Electronic Materials and Wearable Devices, Sun Yat-sen University, Guangzhou, 510275, China

[b] Guangdong Provincial Key Laboratory of Magnetoelectric Physics and Devices, School of Physics, Sun Yat-sen University, Guangzhou, 510275, China

[c] Joint Key Laboratory of the Ministry of Education, Institute of Applied Physics and Materials Engineering, University of Macau, Macau, Taipa, 999078, China

[d] School of Environmental Science and Engineering, Sun Yat-sen University, Guangzhou, 510275, China

E-mail: yongqingcai@um.edu.mo (Y. Cai), yank9@mail.sysu.edu.cn (K. Yan), luohx7@mail.sysu.edu.cn (H. Luo)

**Abstract:**

Topological semimetals with protected surface states mark a new paradigm of research beyond the early landmarks of band-structure engineering, allowing fabrication of efficient catalyst to harness the rich metallic surface states to activate specific chemical processes. Herein, we demonstrate a facile solid-phase method for in-situ doping of Ir at the Os sites in the $Os_3Sn_7$, an alloy with topological states, which significantly improves the photocatalytic performance for the reduction of $CO_2$ to CO and $CH_4$. Experimental evidence combined with theoretical calculations reveal that the nontrivial topological surface states greatly accelerate charge-separation/electron-enrichment and adsorption/activation of $CO_2$ molecules, rendering highly efficient reaction channels to stimulate the formation of *COOH and *CO, as well CHO*. This work shows the promise of achieving high photocatalytic performances with synthesizing topological catalysts and provides hints on the design of novel topological catalysts with superior photoactivity towards the $CO_2$ reduction reaction.

## 1. Introduction

Development of advanced photocatalytic technology, crucial for the conversion of solar-to-chemical energy, is highly demanding for achieving efficient and stable photocatalysts with a high rate of the photochemical reaction [1-7]. The conversion of carbon dioxide ($CO_2$) from fossil fuel combustion into value-added chemicals and fuels has emerged as a promising avenue to mitigate the increasing levels of $CO_2$ in the atmosphere and simultaneously enrich the energy supplies, which has been recognized as a key to reducing carbon emissions and achieving carbon neutrality [8-14]. Amongst various mono-carbon ($C_1$) chemical products converted via photocatalytic $CO_2$ reduction reaction (CRR), such as formates, carbon monoxide (CO), methane, ethane, ethanol, and methanol, CO is the most common, simple and high-value product [15-21]. Comparing with other products, the conversion of $CO_2$ to CO has the advantage of involving only a two-electrons reaction, which however often competes with hydrogen evolution reaction (HER), leading to a low internal quantum efficiency ($IQE_{cr}$) for CRR [8, 22, 23]. In view of this, the development of a highly efficient photocatalyst that is more conducive to catalyzing CRR than HER is the key to solving this problem.

In terms of catalytic properties (e.g., catalytic activity, $IQE_{cr}$, and product selectivity), the surface state of the catalyst usually plays an appreciable role than the bulk state. Such metrics adopted in the reaction largely relies on efficient and robust sampling of the surface states which are prone to uncertainties induced by impurities and lattice imperfection. To this end, topological phase materials, with topologically protected rich surface states, act as unique platform to explore the interactions between surface state, carrier separation/transfer, and surface catalytic reactions [24-27]. In particular, topological semimetals (such as Nodal-line and Weyl semimetals) have emerged as ideal candidate various types of catalysts (photocatalysts and electrocatalysts, etc.) on account of their nontrivial topologically-protected surface states, high electrical conductivity, and suitable carrier concentrations near the Fermi energy ($E_F$) levels [28, 29]. In principle, the surface protected metallic states in topological compounds, well localized at the surface and reminiscent of metallic states of traditional noble metals, can be utilized for activating chemical reaction. For instance, $PtSn_4$ and $Co_3Sn_2S_2$ semimetals have been suggested or experimentally tested as high-performance and stable electrocatalysts for the oxygen evolution reaction (OER) and HER due to the presence of an abundance of the topologically protected surface states [30]. An updated review of semimetal catalysts, ruthenium tin ($Ru_3Sn_7$) has become a rising star in the field of electrocatalytic conversion thanks to its high chemical inertness, suitable topological band structure, low cost, and other advantages [31]. Unfortunately, due to the slow charge dynamics limiting the effective utilization of photogenerated charge carriers, $Ru_3Sn_7$, together with other analogues such as $PtSn_4$ and $Co_3Sn_2S_2$, as a catalyst of the $CO_2$ conversion process still faces challenge of promoting its efficacy.

In view of the above facts, through an ensemble effect, alloying a high-performance active metal with other high-abundance metals is an effective strategy to customize the adsorption of specific reaction intermediates. On this front, the active metal osmium (Os), which is the lowest price among all platinum-group metals (rhodium (Rh), palladium (Pd), platinum (Pt), ruthenium (Ru), iridium (Ir), and Os) and possesses a similar *d*-band structure to Ru, is a promising mixing element [32, 33]. Following this strategy, the development of Os-based materials can be a technological reserve for the noble-metal catalyst systems. In this work, we combine topological surface states and highly conductive bulk phases to prepare a nontrivial semimetal $(Os_{0.5}Ir_{0.5})_3Sn_7$ through a facile electrical arc-melting strategy and high-temperature solid-phase treatment, which can significantly improve the CRR kinetics. The bulk metallic nature of $(Os_{0.5}Ir_{0.5})_3Sn_7$ ensures a fast carrier transport and kinetics. Meanwhile, high carrier concentration and strong carrier transfer between negatively charged Os/Ir atoms and Sn atoms are believed to be essential for the generation of *COOH ,*CO, and CHO* intermediates. This work reveals the photocatalytic performance of dual-atomic alloy photocatalysts with nontrivial topological surface states, which offers promise for the design of novel semimetal catalysts with excellent photoactivity toward CRR.

## 2. Experiment section

### 2.1 Materials and Reagents

Iridium powder (Ir, 99.999%), osmium powder (Os, 99.999%), and tin powder (Sn, 99.999%) were purchased from Shanghai Aladdin Biochemical Technology Co., Ltd. and used without further purification. All chemicals were of analytical grade and were used as received without purification.

### 2.2. Preparation of $Os_3Sn_7$, $Ir_3Sn_7$, and $(Os_{0.5}Ir_{0.5})_3Sn_7$ semimetals

The preparation of all the samples was referred to as a modified arc-melting method and high-temperature solid-phase preparation. First, Ir, Os, and Sn powders were weighed according to the nominal content of the desired compounds. Then, the powder mixture was ground in a glovebox under Ar atmosphere and pressed under a tablet press to form a sheet with a diameter of 6 mm. Finally, $Os_3Sn_7$, $Ir_3Sn_7$, and $(Os_{0.5}Ir_{0.5})_3Sn_7$ were synthesized in the arc-melting furnace at the reaction temperature up to 3000 °C, followed via high-temperature solid-phase treatment (650 °C). In order to more intuitively describe the performance test of the samples, in this work, $Os_3Sn_7$, $Ir_3Sn_7$, and $(Os_{0.5}Ir_{0.5})_3Sn_7$ were denoted as OS, IS, and OIS, respectively.

### 2.3. Photocatalytic CRR measurements

The photocatalytic CRR measurement was conducted in an online trace gas analysis system with a gas chromatography-mass spectrometry (GC-MS, Agilent GC/MS-7000D) and $^1$H nuclear magnetic resonance ($^1$H NMR) spectroscopy, where 3 mL $H_2O$ was injected into the bottom of photocatalytic reactor before sealing the system that ensured the $H_2O$ vapor was saturated during photocatalysis. In detail, 50 mg of photocatalysts, 20 mL of triethanolamine (TEOA) solution, and 100 mL of aqueous acetonitrile (MeCN) solution (MeCN:$H_2O$ = 4:1) were added in the Pyrex glass reaction cell with sonication. It is worth noting that the sample prepared by arc-melting and high-temperature treatment needs to be further ball ground into a fine powder. The reactor system was then filled with pure $CO_2$ gas following complete evacuation. After adsorption equilibrium in a dark environment, the reactor system was put under a simulated light source which was provided by a 300 W Xe lamp (Beijing Perfect light, Microsolar 300, 100 mW·cm$^{-2}$). In addition, the reaction system was controlled at 298 K by circulating water. Finally, the produced gas was analyzed by every 1 h through GC and $^1$H NMR spectroscopy (Bruker AVIII HD 600), respectively. The error bars for gas evolution uncertainty represent one standard deviation based on 3 independent samples.

## 3. Results and discussion

### 3.1. Morphology and crystal structure characterization

Among all the prepared OS, IS, and OIS semimetals, the OIS exhibited the highest photocatalytic activity and was therefore selected as the proof-of-model catalyst. Powder X-ray powder diffraction (PXRD) data and the corresponding Rietveld refinement analysis indicate that a crystal structure of OIS is similar to that of osmium tin ($Os_3Sn_7$, Im-3m, 229) and iridium tin ($Ir_3Sn_7$, Im-3m, 229), with no discernible other crystalline phase observed (Figs. S1a and b, and Fig. 1e). Noteworthily, the intensity of the (330) crystal facet (i.e., $2\theta$ = 40.90°) of OIS is higher than that in the OS and IS databases, suggesting that the (330) crystal plane is conducive to the crystal growth process. Figs. S2 and S3 illustrates the crystal structures of OIS and IS with (100), (001), and (010) crystal facets. The results mean that OIS processes a highly symmetric cubic lattice with a space group of Im-3m (229), and its higher symmetry enables OIS the possibility of being topological. In addition, the ICP-OES content analysis of OIS, OS, and IS samples, and the obtained content of elements is similar to the stoichiometric ratio (Table S1).

To look into the detailed structures, the OIS, OS, and IS semimetals were studied via high angle annular dark-field scanning transmission electron microscopy (HAADF-STEM) (Fig. 1a), where subatomic resolution can be achieved on account of the rather high quality of the prepared crystals. The inset in Fig. 1a is the [3-73] directions of OIS, which matches the STEM image perfectly, indicating higher crystalline properties. The alloying of Os and Ir in the OIS is examined by aberration-corrected HAADF-STEM (AC HAADF-STEM) shown in Fig. 1b. The atomic distribution of OIS, OS, and IS semimetals was confirmed by AC HAADF-STEM, where the green and red atoms in Fig. 1b represent Os and Ir, respectively, and the gray atoms represent Sn, proving that Os and Ir occupied the same atomic sites. The Fourier transform (FFT)-filtered atomic resolution images (Figs. 1f and g) further display that Os and Ir atoms occupy the identical sites, and that the ratio of Os to Ir is constant on each atomic column [34]. One can observe three brightness levels in Figs. 1b and f. Nevertheless, this does not originate from a third element but due to the superposition of Os/Ir and Sn atom columns at specific points [31]. Several typical lattice distances exhibited in Fig. S3a were measured and are shown in Fig. S3b–f, where the measured values of 6.59 and 4.68 Å correspond to the (110) and (200) faces of OIS, respectively. The selected area electron diffraction (SAED) patterns of OIS and OS are displayed in the insert of Fig. 1a, respectively, which can be indexed along the [3-73] directions, respectively. Subsequently, the simulation plots of the SAED pattern are also given in this paper, in good concert with the experimental results (Fig. 1c). From the HAADF-STEM and elemental mapping images (Fig. 1d), it can be obviusly seen that Ir, Os, and Sn elements are uniformly distributed in the OIS. Likewise, a series of parallel characterizations of OS semimetals were tested and are shown in Fig. S4. The above detailed analytical results prove that OIS, OS, and IS semimetals, prepared in this work, have high purity and crystallinity, and can be used as ideal control samples to highlight the effect of topological nontrivial bands on the catalytic activity of CRR.

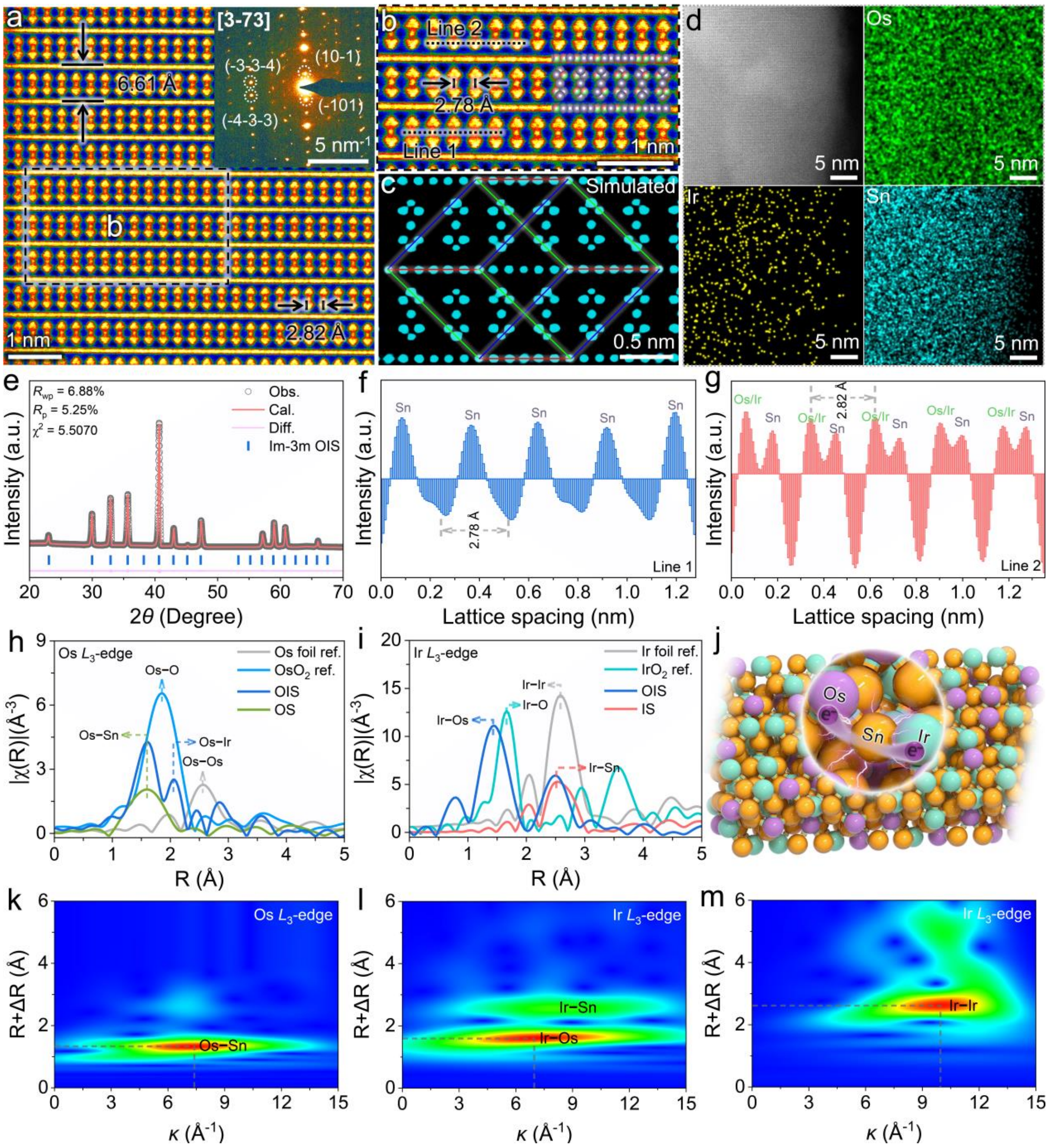

**Fig. 1.** (a) HAADF-STEM image taken in the [3-73] direction of OIS and the inset of SEAD pattern. (b) Magnified HAADF-STEM image taken from the corresponding area in (a) (top), and (c) the simulated microscopic image of OIS from [3-73] direction (bottom). (d) EDS mapping of OIS nanocrystal. (e) Rietveld refinement of XRD data for OIS. Intensity line profiles obtained from the area marked by black rectangles in (f) panel (b: line 1) and (g) panel (b: line 2). (h) Os and (i) Ir $L_3$-edge EXAFS spectra in $R$ space. (j) Schematic illustration of the optimized CRR process induced by the OIS. WT for (k) Os $L_3$-edge of OIS and (l) Ir $L_3$-edge of OIS and (m) Ir foil ref.

3.2 Electronic structure investigation

The X-ray absorption near-edge structure (XANES) and extended X-ray absorption fine structure (EXAFS) were performed to determine the precise electronic structures and detailed local coordination environments of OIS, OS, and IS semimetals on the atomic level (Figs. 1h and i, S5 and S6). As shown in the Os and Ir $L_3$-edges XANES spectra (Figs. S5a and b), the absorption edge energies of OIS, OS, and IS are close to the reference materials of Os and Ir foils, whereas the intensity of the white line (i.e., orbital transition of $2p \rightarrow 5d$) at ~10875 and 11215 eV is much weaker than that of $OsO_2$ and $IrO_2$, respectively, which implies that Os and Ir elements in the OIS, OS, and IS semimetals are predominantly in the metallic state [35, 36]. Compared with Os and Ir foils, the absorption edge position of Os $L_3$-edge and Ir $L_3$-edge has a negative shift after interacting with Sn atoms, suggesting the transfer of electron density from Sn to Os and Ir atoms. The intensity of the white line in XANES at the Os $L_3$-edge and Ir $L_3$-edge of OIS is slightly lower than those of the OS and IS, which may be on account of the alloying with Sn atoms resulting in more filling of the Os and Ir $5d$ bands. The above results imply the moderate electron transfer in the OIS, compared with the OS and IS, respectively [36, 37]. X-ray photoelectron spectroscopy (XPS) survey spectrum further discloses that the Os, Ir, and Sn elements are dominant in OS and IS (Fig. S1c−f). In the Os $4f$ XPS spectra, the Os $4f_{7/2}$ and Os $4f_{5/2}$ peaks of OS are at 50.8 and 53.6 eV, respectively, the peak positions of which are higher than those reported for Os nanoparticles (50.3 and 53.1 eV, respectively) (Fig. S1d), the oxidation state of Os elemental is close to neutral [38]. One pair of doublets at 62.6 and 59.6 eV ($Ir^{3+}$ $4f_{5/2}$ and $Ir^{3+}$ $4f_{7/2}$) is ascribed to iridium oxide. The other pair of doublets at 62.1 and 58.9 eV ($Ir^0$ $4f_{5/2}$ and $Ir^0$ $4f_{7/2}$) corresponds to metal Ir species (Fig. S1e) [39, 40]. As expected, the binding energy of the Sn $3d$ peak is 484.8 eV, which can be attributed to $Sn^0$ $3d_{5/2}$ [41]. In addition, both of the main peaks are composed of $Sn^{4+}$, which is mainly due to the surface reoxidation by air (Fig. S1f). Notably, the proportion of $Os^0$ and $Ir^0$ in OIS is determined to be 65.24% and 64.35%, higher than those of 42.18% and 55.62% in OS and IS, respectively, (Table S1), demonstrating the valance states are closer to 0 and increased in the electron-cloud density of Os and Ir atoms. Interestingly, Sn 3*d* XPS spectra represents that the double-state splitting energy of OIS is decreased to 8.50 eV compared with 8.53 eV for OS and IS, respectively (Fig. 2a). The decrease splitting energy of OIS confirms the reduce of electron-cloud density around the Sn atoms, which is indicative of the donor behavior of the Sn atoms, largely from its *d* orbitals.

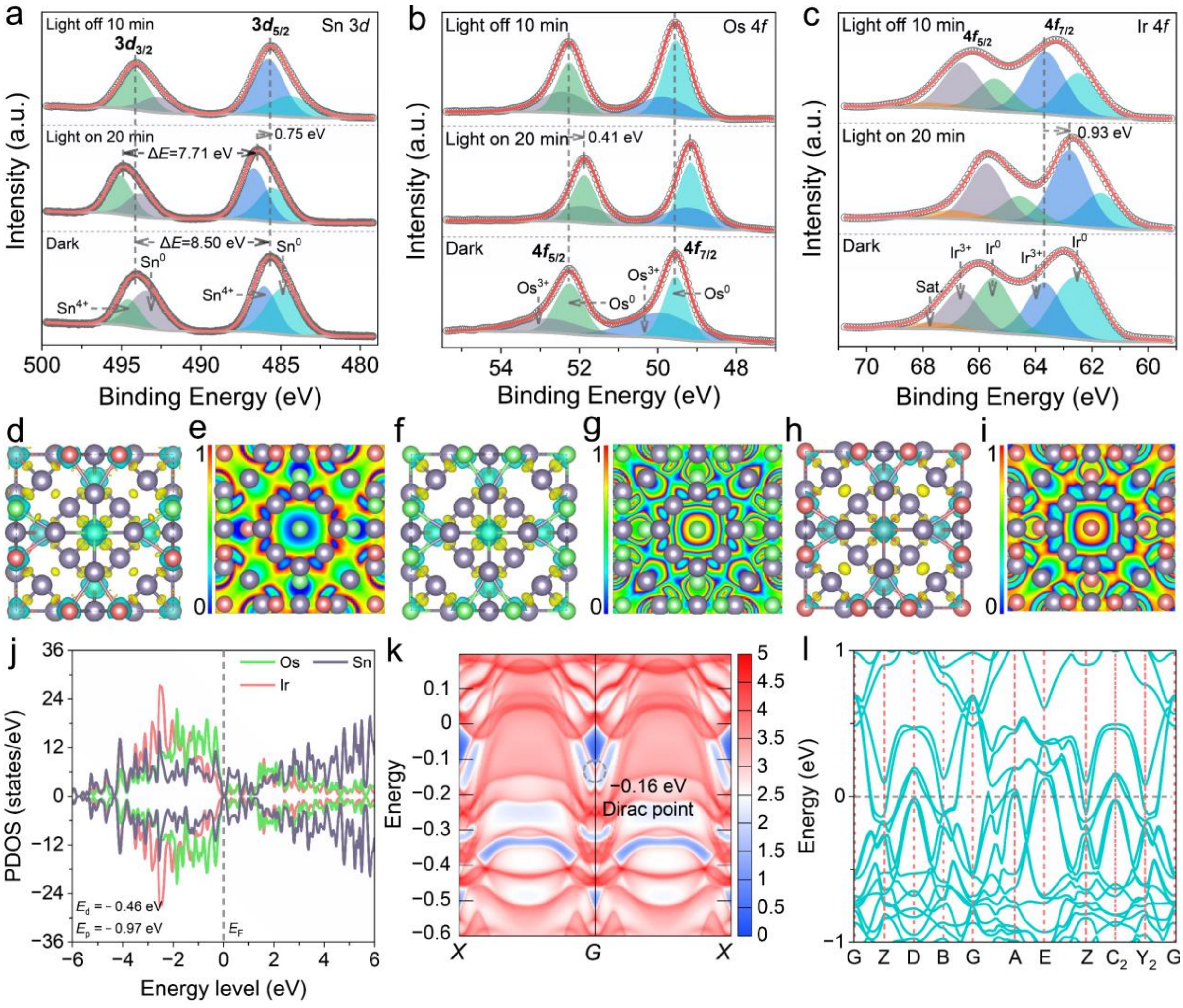

**Fig. 2.** In-situ XPS (a) Sn 3*d*, (b) Os 4*f*, and (c) Ir 4*f* spectra of the OIS. Charge Density Difference of on the (d) OIS, (f) OS, and (h) IS (charge accumulation is in cyan and depletion in yellow) with dipole moment correction, respectively. ELF diagrams the (e) OIS, (g) OS, and (i) IS, respectively (The gray, red, and green spheres represent Sn, Ir, and Os atoms). (j) The calculated DOS of OIS. (k) DOS of the (001) facet of OIS (view of the TSS in black circle). (l) The enlarged perspective of bulk electronic band structure of the OIS.

Indeed, the elaborate design of active sites is central to optimizing the adsorption energetics of the reactants and intermediates by tuning the electron distribution of the catalysts. The charge transfer between Os, Ir, and Sn atoms in the OIS semimetal is further investigated by in-situ XPS. After 20 min of illumination, the peaks corresponding to the binding energy of Ir 4*f* and Os 4*f* shift to the right via 0.93 and 0.41 eV, respectively, and then recover to the situation of "dark-state" after 10 min of light-off, signifying that the electronic gain of Ir atoms upon light illumination (Figs. 2b and c). The binding energy change of Sn 3*d* further confirms the direction of electron migration (Figs. 2c). At the same time, the bistatic splitting energy of the Sn 3*d* decreases by 0.77 eV under light irradiation in OIS, indicating that photoelectrons are transferred from Sn atoms to Os and Ir atoms. In addition, the presence of Os−Sn−Ir asymmetric species with strong interactions leads to the migration of electrons from Sn atoms to Os and Ir atoms to form donor-acceptor structures at the atomic scale, which reduces Gibbs free energy change ($\Delta G$) of the conversion process from $CO_2$ to *COOH, and provides the necessary force for the improvement of the catalytic efficiency [42]. The transfer of electrons from Sn atoms to Os and Ir atoms is further demonstrated via density functional theory (DFT) calculations and Bader charge analysis, with negative charges accumulating mainly in the Os and Ir atomic regions, as described in Figs. 2d, f, and h [43].

In view of the promoted activity of OIS for CO and $CH_4$ production, the fundamental electronic properties and electron localization function (ELF) diagrams are further explored and analyzed. The electrons locate around the hollow position between the Os/Ir on the pure OS/IS (Figs. 2g and i), reflecting the strong activity of the hollow site for the Os/Ir atoms [44]. Moreover, for Os and Ir atoms in OIS, the electron accumulation regions extending along the Os−Sn and Ir−Sn bonding directions can be clearly seen (Fig. 2e). Simultaneously, the electron localization around the hollow site becomes weaker while the electrons begin to accumulate more to the Os and Ir atoms from OIS. For the pure IS (Fig. 2i), the ELF plots display the electron localization around the Ir atoms rather than at the hollow region. Since the electronegativity of Ir is larger than the Os, the charge of the Ir atoms can become more negative in the OIS. That is to say, the Ir atoms in OIS reduces the activity of Os atoms, which in turn strengthens the hydrogenation reaction [45]. In comparison with pure OS and IS, the adsorption activities of H atoms are weakened by small amounts of Ir atoms in OIS and then the catalytic property of $CO_2$ reduction to $CH_4$ and CO is enhanced at the OIS.

The density of states (DOS) and corresponding bulk electronic band structures of OIS, OS, and IS semimetals without spin-orbit coupling (SOC) reveal the formation of orbital hybridization around the $E_F$ levels of Os-5*d*, Ir-5*d*, and Sn-3*d*, as presented in Figs. 2j, S7a and b. The few crossings and low DOS at the $E_F$ levels indicate the semi-metallic properties of the material. Interestingly, the DOS analysis further means that the OIS possesses a *d*-band center closer to the $E_F$ levels, where there is an abundance of electrons available for binding to reaction intermediates (e.g., *$CO_2$, *COOH, *CO, and CHO*; the asterisk denotes the catalytically active sites). Of note, it was found the existence of topologically protected surface states in the bandgap at the *G* point of OIS (001) surface (Dirac cone, −0.16 eV) (Fig. 2k), and its surface state energy level is near the $E_F$ (Fig. 2l), which implies that the surface states can provide additional electrons and promote the transfer of electrons from the OIS (001) surface to the adsorbed $CO_2$ molecules. In addition, quasi-linear dispersion relation is observed in the surface state, indicating high electron mobility, which remarkably is conducive to CRR [46, 47]. To further elucidate the role of topological surface states, the surface DOS of topologically trivial

(110) crystal facet family in OIS, which are selected according to the XRD results, were also calculated. The corresponding relaxed structures for DFT calculations can be found in Figs. S7c and d. The surface DOS of OIS (001) and (110) facets are presented in Figs. S7e and f, respectively, where no topological surface states can be found in (110) facets. Therefore, it is generally conjectured that the topological OIS (001) facets would exhibit marvelous activity for CRR, which can be verified by the following calculations and experiments.

3.3. Photocatalytic CRR performance

The photocatalytic performance of OIS, OS, and IS semimetals under visible light irradiation (≥ 420 nm) was examined by an independently constructed photocatalytic CRR device, as shown in Figs. 3, S8a and b. Initial control experiments were conducted individually without photocatalyst, light, or $CO_2$. The results indicate that the photocatalyst and light source were essential for $CO_2$ reduction (Figs. S9a and b), and only little amounts of $H_2$ and $O_2$ were produced in the absence of $CO_2$. OS and IS were almost incapable of photocatalytic reduction of $CO_2$, and the gas product yields of CO and $H_2$ over OS (IS) are 25.50 (8.65) $\mu mol \cdot g^{-1} \cdot h^{-1}$ and 11.15 (15.54) $\mu mol \cdot g^{-1} \cdot h^{-1}$, respectively (Fig. 3a). By contrast, the CO photocatalytic yield (119.30 $\mu mol \cdot g^{-1} \cdot h^{-1}$) of OIS is 5 and 14 times higher than that of OS and IS, respectively. Additionally, the gas product yield of $CH_4$ over OIS is 4.3 $\mu mol \cdot g^{-1} \cdot h^{-1}$ (Fig. 3a), and OS and IS are nearly photo-catalytically incapable toward $CO_2$ reduction to $CH_4$. The gas product yields of CO and $H_2$ over OIS are 33.1 and 18.2 $\mu mol \cdot g^{-1} \cdot h^{-1}$ without TEOA as a sacrificial agent, respectively, and product-based selectivity of CO is achieved at 64.5% (Fig. S9d). It is worth to mention here that the CO production yield of OIS, likely to be a new record in terms of CO production, is higher than those values declared for metal-, photosensitizer-, and sacrificial agent-free photocatalytic systems up to now (Table S2). $CO_2$ adsorption isotherms reveal that OIS possesses the highest $CO_2$ uptake capacity, which is the prerequisite step for triggering subsequent CRR (Fig. S8c). An electron-based selectivity and product-based selectivity of CO over the OIS are achieved at 86.7%, which is 1.25- and 2.43-fold higher than that of the OS and IS (Figs. 3b and c). As the photocatalyst mass loading increases, the CRR performance of OIS is promoted further and reached up to 5.97 μmol·h$^{-1}$ for 50 mg (Fig. S9c), with a nonlinear increase of mass-normalized CRR with the photocatalyst mass, due to an enhanced light scattering caused by a large amount of photocatalyst [48]. As shown in Figs. 3d and e, there is no apparent decline after nine runs of photocatalytic testing (retains 86.5% of its original catalytic activity), demonstrating decent stability of OIS. Furthermore, no apparent changes were observed in the AC HAADF-STEM, XRD, XANES, and XPS (Fig. S10) characterizations of OIS after the duration of 45 h, indicating its excellent structural stability. Markedly, the control experiments in different conditions (Fig. S9d) confirm that the detected products are indeed derived from the reaction between $CO_2$ and water ($H_2O$) catalyzed by the samples. The carbon source was verified through the $^{13}C$ isotopic composition of the gas products, as presented in Figs. 3f, S8d and e. The $^{13}C$ labeled isotope experiments further imply that the reduction products ($CH_4$ and CO) originated from $CO_2$ rather than from decomposition of the materials or other residual organics during the catalytic reaction. The $IQE_{cr}$ of OIS at 380, 400, 420, 440, 460, 480, 500, and 520 nm were tested and calculated at 0.218%, 0.199%, 0.194%, 0.189%, 0.166%, 0.165%, 0.09%, and 0.04%, respectively (Table S3) [11, 23, 49, 50]. These results deficient that the photocatalytic CRR activity can be significantly promoted by the synergy between Os and Ir in a nontrivial topological semimetal OIS.

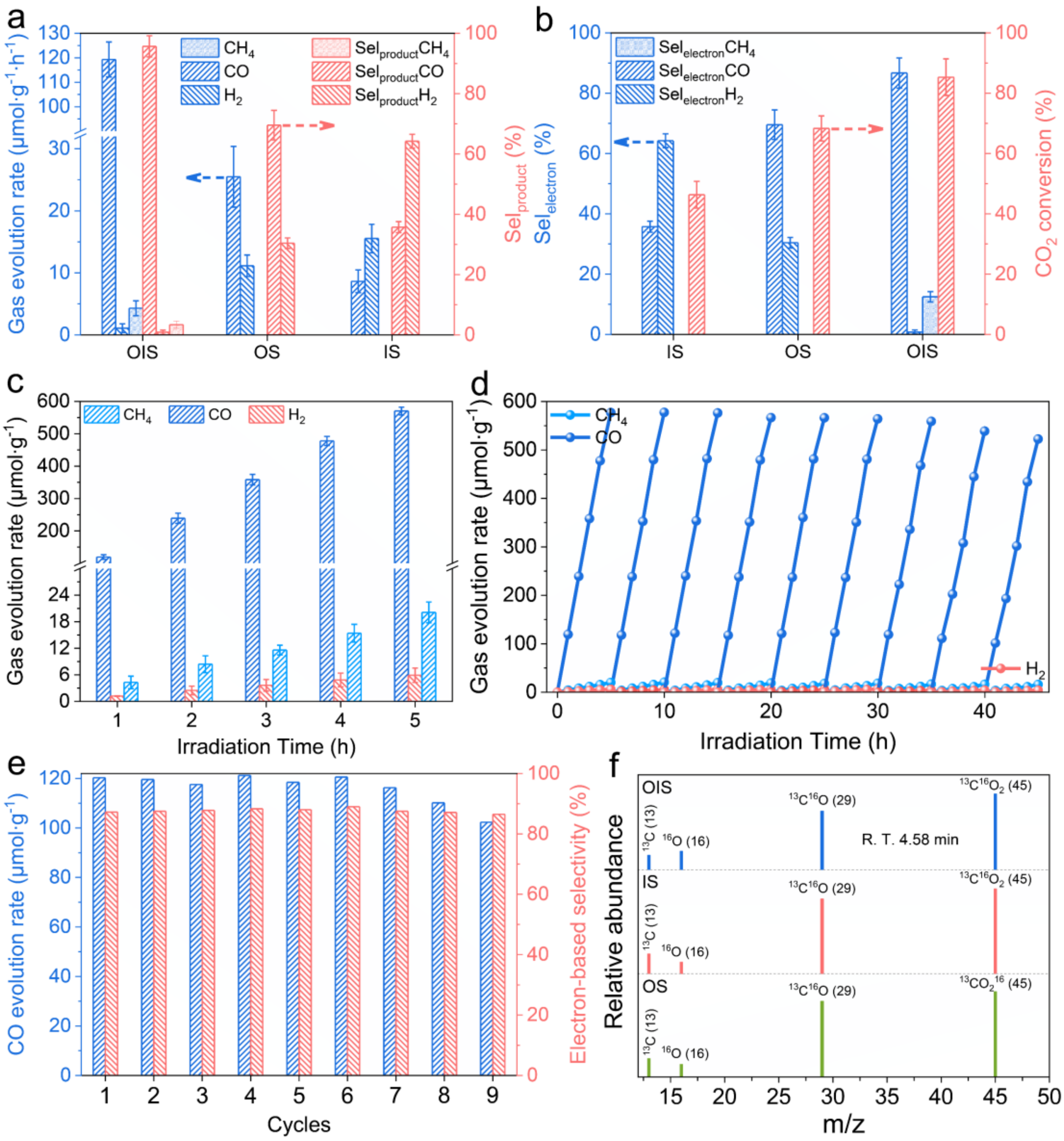

**Fig. 3.** (a) Product formation rate and selectivity of samples in MeCN solution. (b) Electron-based selectivity and $CO_2$ conversion on samples in MeCN solution. (c) Photocatalytic product evolution as a function of light irradiation times on OIS in MeCN solution. (d) Gas evolution amounts as a function of light irradiation time for OIS over nine cycling tests in MeCN solution. (e) Product formation rate and electron-based selectivity on OIS over nine cycling tests in MeCN solution. (f) GC-MS of $^{13}C^{16}O$, $^{13}C^{16}O_2$, $^{13}C$, and $^{16}O$ produced from the photocatalytic reduction of $^{13}CO_2$ for different photocatalytic system.

3.4. Photoelectric performance analysis

To understand the significantly enhanced photocatalytic activity of OIS, ultraviolet-visible (UV-vis) diffuse reflectance spectra (DRS) were used to evaluate the optical absorption properties of OIS. An absorption edge at around 455 and 532 nm is observed for IS and OIS. The absorption edge of OIS is slightly larger than that of OS and IS, respectively, verifying a blue-shift absorption edge of OS and IS (Fig. S11a), which are similar to the photoelectrochemical test results of OIS, OS, and IS semimetals (Figs. 4a and S11b) [51]. Advanced characterization techniques, including steady-state photoluminescence (PL), ultrafast femtosecond transient absorption (fs-TA), and Hall effect measurements, were systematically employed to gain insights into the recombination and lifetime of charge carriers of the prepared semimetals. As shown in Fig. S11c, all the OIS, OS, and IS semimetals show a broad PL peak at around 432 nm, and PL intensity of IS is significantly higher than that of OIS and OS, respectively [52]. Despite suffering from the appreciable difference in PL intensity, the TRPL curves declare almost the same decaying trend in Fig. S11d. The TRPL data curves fit well with the stretching exponential function, and the specific fitting parameters are shown in Table S4, signifying that the average lifetimes of IS, OS, and OIS for the stretching exponential decays are 1.8344, 1.7447, and 1.7327 ns, respectively. To gain further insight into the involvement of carrier transfer properties in photocatalysts, the real-time dynamics of the photogenerated carriers of OS and OIS were studied using fs-TA spectra. First, we tested the transient absorption spectra of the OS molecules as controls at different pump-probe delay times (Figs. 4d and e). The redshifts of the GSB peaks over time can be observed in both Fig. 4d. This is because OS possess a broad size distribution, in which the smaller OS has the faster exciton annihilation. Upon excitation by a pump pulse with a wavelength of 320 nm, the TA spectrum of OIS reveals an obvious negative peak at around 360 nm, which belongs to the ground-state bleach (GSB), reflecting the relaxation of the excited state (Fig. 4b), which is attributed to the charge transfer between Os/Ir and Sn atoms rather than other process occurring in OIS. The TA spectra of OIS also show distinct a positive absorption bands, which is assigned, respectively, to the excited-state absorption (ESA) signal. Fig. 4d displays a pseudo-color plot of the TA spectra of OIS with probe wavelength and delay time. Moreover, the contour plots of a subset of TA spectra taken at different probe delays show that OIS has that a broad positive absorption in the wavelength range of 390 to 450 nm (Figs. 4c and e). Intriguingly, the blueshifts of the GSB peaks over time can be observed in Fig. 4b. This is due to the introduction of Ir in OS leading to slower exciton annihilation. In the presence of lactic acid (LA, a hole scavenger) or potassium dichromate ($K_2Cr_2O_7$, an electron scavenger), quenching experiments were conducted to analyze the decay kinetics and determine the contribution of photogenerated charge carriers to the GSB signal in depth. As displayed in Figs. 4g and h, the GSB signal of OIS-LA shows a recovery process similar to that of pristine OIS, whereas the signal of OIS-$K_2Cr_2O_7$ is totally quenched. Such spectral features indicate that electrons instead of holes contribute the GSB signal [53-55]. Specifically, OIS exhibits a faster GSB decay rate than OS, agreeing with the Hall effect measurements. The Hall effect measurement further evidences the longer carrier lifetime and stronger positive absorption of samples. The Hall effect measurement at different temperatures indicated that the semi-metallic nature of OIS with a carrier concentration of $9.33\times10^{26}$ $m^{-3}$ and an estimated carrier mobility of $2.36\times10^{-4}$ $m^2\cdot V^{-1}\cdot s^{-1}$ (Figs. 4i and S12). Meanwhile, OS has a carrier concentration of $7.99\times10^{26}$ $m^{-3}$ and a carrier mobility of $1.12\times10^{-4}$ $m^2\cdot V^{-1}\cdot s^{-1}$, respectively. The carrier diffusion lengths ($L_D$) were estimated to be in the range of 0.082~0.26 μm for OIS, 0.058~0.18 μm for OS, and 0.056~0.13 μm for IS, respectively (Table 1) [54]. The improved efficiencies of charge-separation and transfer/electron-enrichment mean significant accessibility and higher utilization of photogenerated carriers, resulting in potential photocatalytic performance advantages.

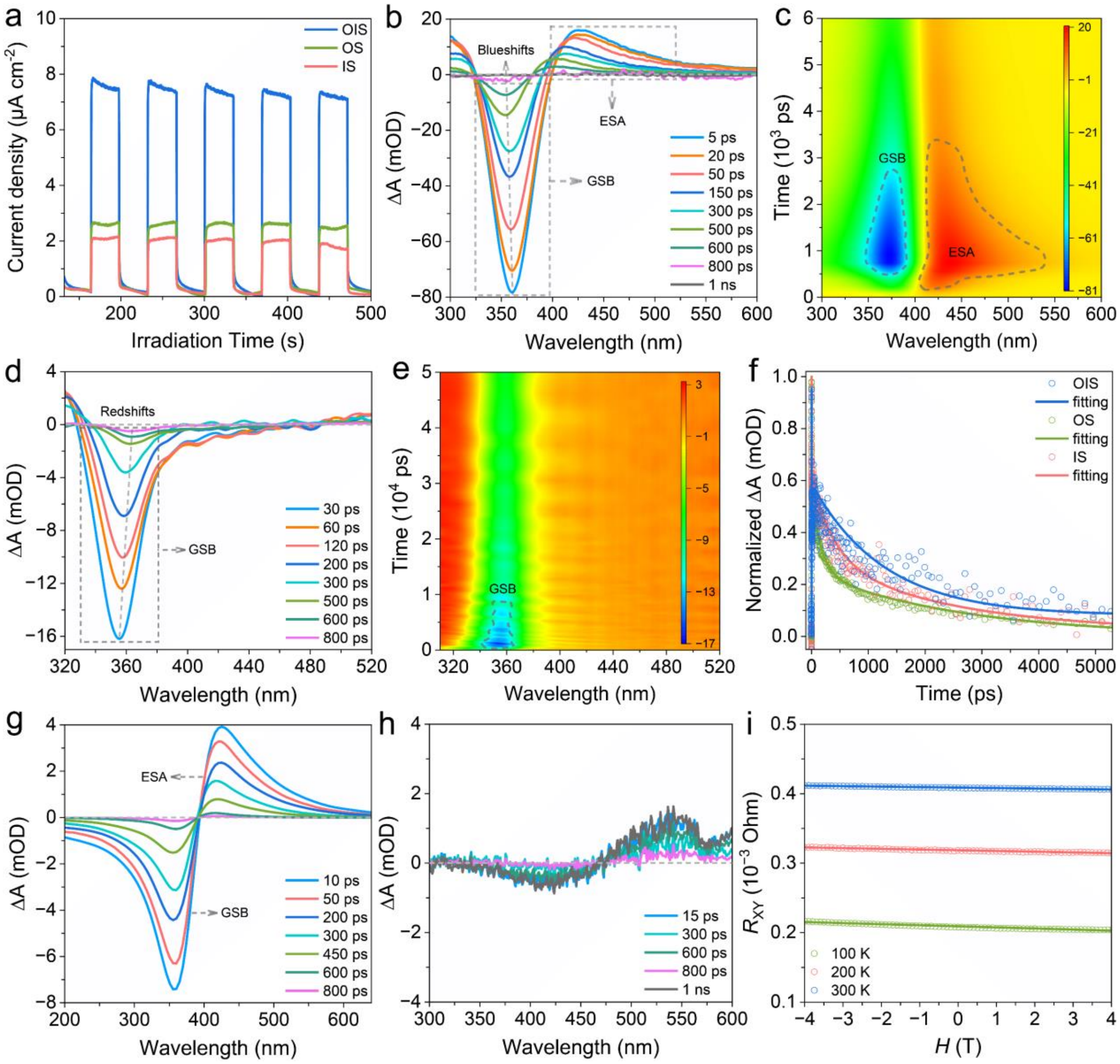

**Fig. 4.** (a) Transient photocurrents (1.0 M KOH) of the OIS, OS, and IS. TA spectra of the (b) OIS and (d) OS at different delays time. 2D pseudo-color images of the (c) OIS and (e) OS in ethanol solution after the excitation with a 320 nm laser pulse. (f) Comparison of decay kinetics and fitting lines for the OIS, OS, and IS. TA spectra of the OIS in the presence of (g) lactic acid as hole scavengers and (h) $K_2Cr_2O_7$ as electron scavengers. (i) Hall resistivity versus magnetic field of the OIS at different temperature.

### 3.5. Reaction mechanism analysis

To provide in-depth insights into the CRR process occurring in the semimetals, we have used a strategy combining of in-situ near ambient pressure XPS (NAP-XPS), in-situ diffuse reflectance infrared Fourier transform spectroscopy (DRIFTS) measurements, and DFT calculations. As shown in Fig. 5a, the C 1*s* spectrum displays an obvious peak at around 284.7 eV in ultra-high vacuum, well assigned to the adventitious carbon on the photocatalyst [56]. Upon light illumination, the extra C 1*s* peaks of surface C−O (286.1 eV) and O=C−O (288.9 eV) species appear and increase with time evolution [57]. Strikingly, OIS and OS semimetals exhibit a discrepancy with respect to the content of O=C−O and C−O (Fig. 5d). The ratio between O=C−O/C−O is ca. 0.34, and the content of O=C−O is as high as 6.34% over OIS after 60 min of reaction, while the ratio decreases to 0.85 with 13.28% of O=C−O can be observed over OS even after 60 min of reaction, indicating that the introduction of Ir atom in OS can promote the conversion of *COOH to *CO and the O=C−O species is likely attributed to the formation of CO [58]. Furthermore, in the O 1*s* spectrum, a major peak is observed at 531.4 eV, which corresponds to the oxygen states of O−H species (Figs. 5b and e). Two new contributions, under the light irradiation, are produced at 532.8 and 530.2 eV for the configuration of C=O and C−O, respectively, and increased in intensity with the increase of irradiation time [59]. By collecting the O 1*s* spectrum in NAP-XPS test, it is also verified the efficient photocatalysis of $CO_2$ reduction. Transmittance peaks are monitored at 1558 for OIS (Fig. 5c) attributable to the *COOH group that is generally regarded as the crucial intermediate in the CRR to CO [54, 60]. Moreover, intensities of the peaks of the above intermediates increased with increasing irradiation time from 0 to 60 min, further implying that the *COOH group is an intermediate during the process of $CO_2$ photoreduction to CO. The emerging peak at 1250 $cm^{-1}$ for OIS (1248 $cm^{-1}$ for OS, Fig. 5f) is attributed to the carboxylate ($*CO_2^-$) vibration, which facilitates the formation of *COOH [54, 61]. The appearance of transmittance band at 1422 $cm^{-1}$ for OIS, respectively, are characteristic of the symmetric stretching of $*HCO_3^-$, indicating that $CO_2$ and $H_2O$ molecules were co-adsorbed onto the OIS [62, 63]. The transmittance bands at 1525 $cm^{-1}$, 1454 $cm^{-1}$, and 1373 $cm^{-1}$ for OIS (1373 $cm^{-1}$ for OS) are attributed to the formation of monodentate carbonate (m-$CO_3^{2-}$) groups, and the bands located at 1397 $cm^{-1}$ and 1331 $cm^{-1}$ for OIS (1396 $cm^{-1}$ and 1332 $cm^{-1}$ for OS) are assigned to the bidentate carbonate (b-$CO_3^{2-}$) groups, which were most likely derived from the reaction between $CO_2$ and $H_2O$ molecules, disclosing that the abundant intermediates could be generated on the surface of OIS [54, 63, 64]. Interestingly, the methyl species ($CH_3$*) character band located at 1470 $cm^{-1}$ and 1280 $cm^{-1}$ were observed, which corresponds to the crucial intermediate of $CH_4$ product [23, 65]. In addition, a small amount of $CH_3O$* (1180 $cm^{-1}$), HCHO* (1180 $cm^{-1}$) is detected, resulting from the intermediates to the production of trace methane [66]. The above observed intermediates confirm the highly efficient photocatalytic reduction of $CO_2$ to CO and $CH_4$. In concert with the spectroscopic and catalytic analytical investigations, the possible reaction pathways of $CO_2$ to CO and $CH_4$ were proposed to be: $CO_2$(g) → *$CO_2$ → *COOH → *CO → CO(g) [67, 68] and *CO → CHO* → $CH_2O$* → $CH_3O$* → $CH_4$(g) [69, 70].

Following the above CRR reaction pathway specifically on the OIS, OS, and IS semimetals surface (Figs. 6a–c), we calculated the $\Delta G$ for the potential-limiting step ($CO_2$ to *COOH) on OIS structure to be 0.83 eV, much lower than that on the OS (1.13 eV) and IS (1.41 eV) (Fig. 6f). The CRR performance on IS (001), OS (001), and OIS (001) surfaces were also calculated in detail. The $\Delta G$ for $CO_2$ to *COOH and *COOH to *CO on IS (001) is 1.05 and −0.66 eV, respectively, which makes it difficult for CO to dissociate from the IS (001) surface [71]. The $\Delta G$ for $CO_2$ to *COOH is about 0.86 and 0.57 eV on OS (001) and OIS (001), respectively, illustrating that $CO_2$ adsorption and activation on OIS (001) surface is more favorable for CRR. We further constructed OIS (110) to reveal the role of the topologically nontrivial (001) facet of OIS in catalyzing CRR. Accordingly, the equivalent (110) facet of OIS is constructed and analyzed as a comparison with the topological (001) facet [31]. The $\Delta G$ of topologically trivial planes such as OIS (110) is not as satisfactory as in OIS (001), demonstrating improvement strong dependence on the surface index and the topological surface states on the CRR activity. Noticeably, the lower energy barrier of *CO → CO on OIS suggests the easier desorption of CO molecules from Ir sites (Fig. S13), indicating that presence of OIS plays a positive role in enhancing the selectivity of CO during the photocatalytic CRR [67].

Then, the Δ*G* for $CO_2$ to *COOH on the OS (110) and IS (110) surface is 1.48 and 1.92 eV (Fig. S14a), respectively, much larger than that of OIS (110) (1.24 eV), further denoting that the topological OIS (001) facet is superior for CRR, in agreement with electronic structure analysis results. To gain further insight into the involvement of topological surface states, the DOS of the (001) facet of OIS after adsorption of $CO_2$ molecules was calculated in detail. The surface card is set to be the plane (110), and the calculation is performed by taking 101 slices of one reciprocal vector. The plane (110) is shown in Fig. 6d, and in that plane, the path taken for the edge state spectrum calculation is $X \rightarrow G \rightarrow X$. A possible Dirac cone is observable at the *G* point near energy −0.18 eV in OIS (001) surface. It can be clearly that there are topologically protected surface states in the band gap at the *G* point of OIS (001) surface (Figs. 6d and e), which means that the surface states can provide additional electrons and facilitate the transfer of electrons from the OIS (001) surface to the adsorbed $CO_2$ molecules. On OIS (001) facet, the formation of CHO* from *CO exhibits the highest Δ*G* with a value of 0.83 eV (Fig. 6g), which is energetically more favorable compared with that of OS and IS. According to previous literature, the CHO* formed $CH_3O$* through several proton-electron coupling processes is the selectivity-determining intermediate step [42, 72]. However, the continuous hydrogenation from CHO* to $CH_2O$* on OIS is endothermic, which is less favorable energetically. Thus, both experimental data and theoretical calculations provide compelling evidence that the Os−Ir dual-sites significantly enhanced the performance of $CO_2$ reduction to CO and $CH_4$.

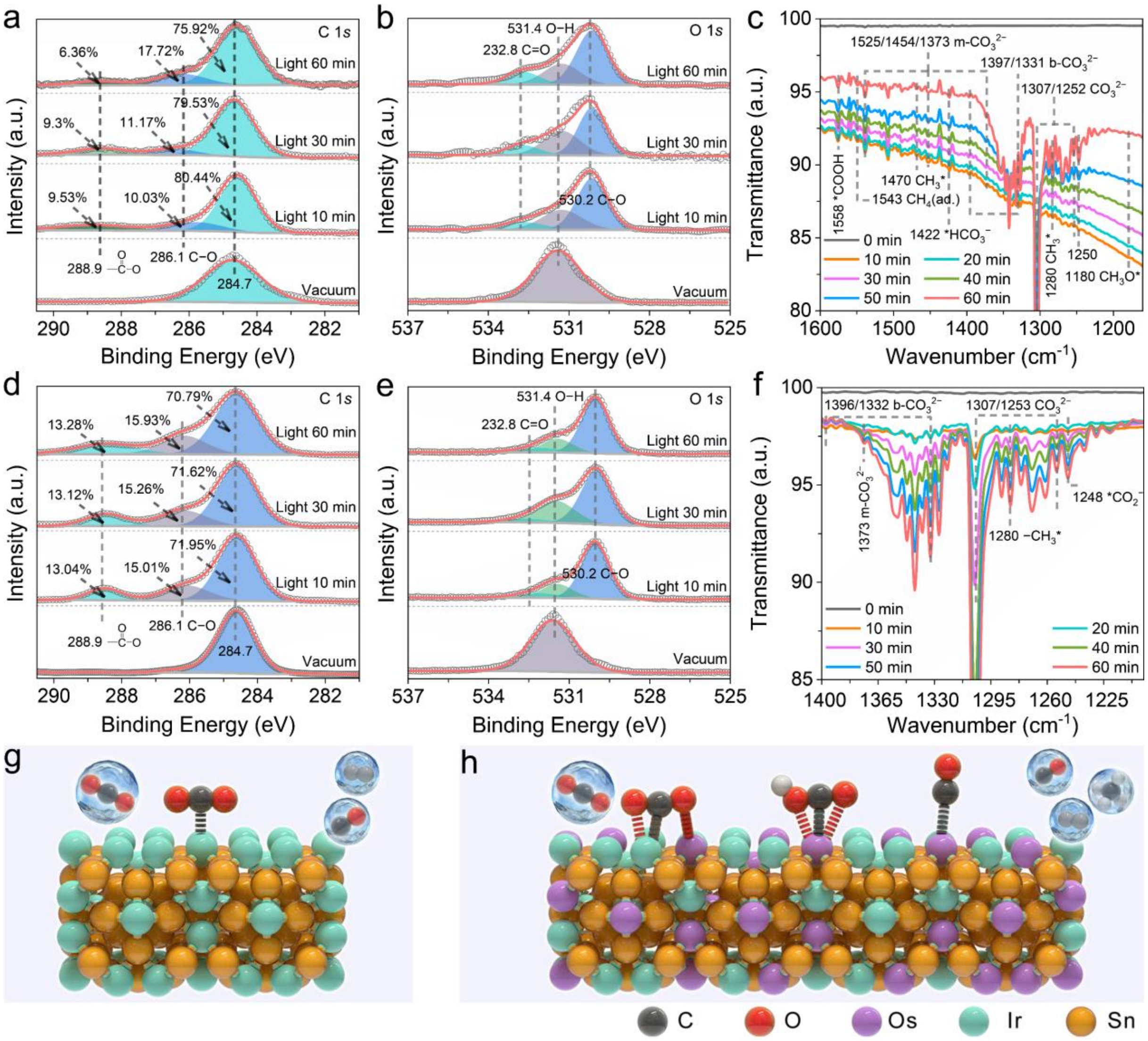

**Fig. 5.** In-situ NAP-XPS results of (a) C 1*s* and (b) O 1*s* spectra over the OIS. (c) In-situ DRIFTS spectra for light-driven $CO_2$ conversion over the OIS. In-situ NAP-XPS results of (d) C 1*s* and (e) O 1*s* spectra over the OS. (f) In-situ DRIFTS spectra for light-driven $CO_2$ conversion over the OS. Reaction pathway schemes for the (g) OS and (h) OIS.

To understand the underlying mechanism of the protonation of $CO_2$, DFT calculations were performed and reveal that the electron cloud around Os and Ir in OIS (Figs. 6i and j) is much more enriched than that in the OS (Fig. S15a–d) and IS (Fig. S15e–h), while is also supported by the electrostatic potential (Figs. 6k, S15i and k) and 2D electron localization function. The localized electronic structure after the formation of *CO and *COOH was simulated, as shown in Figs. 6l, S15f and l. The calculated charge densities of OIS reveal the electron delocalization around Os/Ir and Sn atoms after creating nontrivial topological surface states. The delocalization and overlapping of electron clouds of Os and Ir atoms make the surface metal sites more active for catalytic reactions [73]. In concert with the theoretical calculations and structural, as well catalytic catalysis, OIS with nontrivial topological surface states is rich in electrons close to the $E_F$ levels for adsorption and activation of $CO_2$ molecules, disclosing a greatly enhanced catalytic effect for CRR to CO [74]. Furthermore, HER is known to be a dominant competing reaction for CRR, and the competitive adsorption between *CO and *H on the IS, OS, and OIS semimetals was further examined to study the selectivity of CRR [71, 75]. As shown in Figs. 6c and h, the binding free energies of *H on Os−Sn−Ir dual-sites of OIS are relatively positive ($G_{Os-*H}$ = 0.29 eV, $G_{Sn-*H}$ = 0.47 eV, and $G_{Ir-*H}$ = 0.37 eV), in stark contrast to the negative $G_{*CO}$ (0.22 eV), indicating that Os−Sn−Ir dual-sites can selectively adsorb *CO while retarding *H adsorption [76, 77], thus inhibiting HER and accordingly an enhanced CRR selectivity. Interestingly, as shown in Figs. S14c and d, the $G_{Ir-*H}$ of IS is 0.11 eV and represents a stronger adsorption, in sharp contrast to the positive $G_{*CO}$ (0.38 eV) in favor of HER rather than CRR, which is highly agreement with the experimental results.

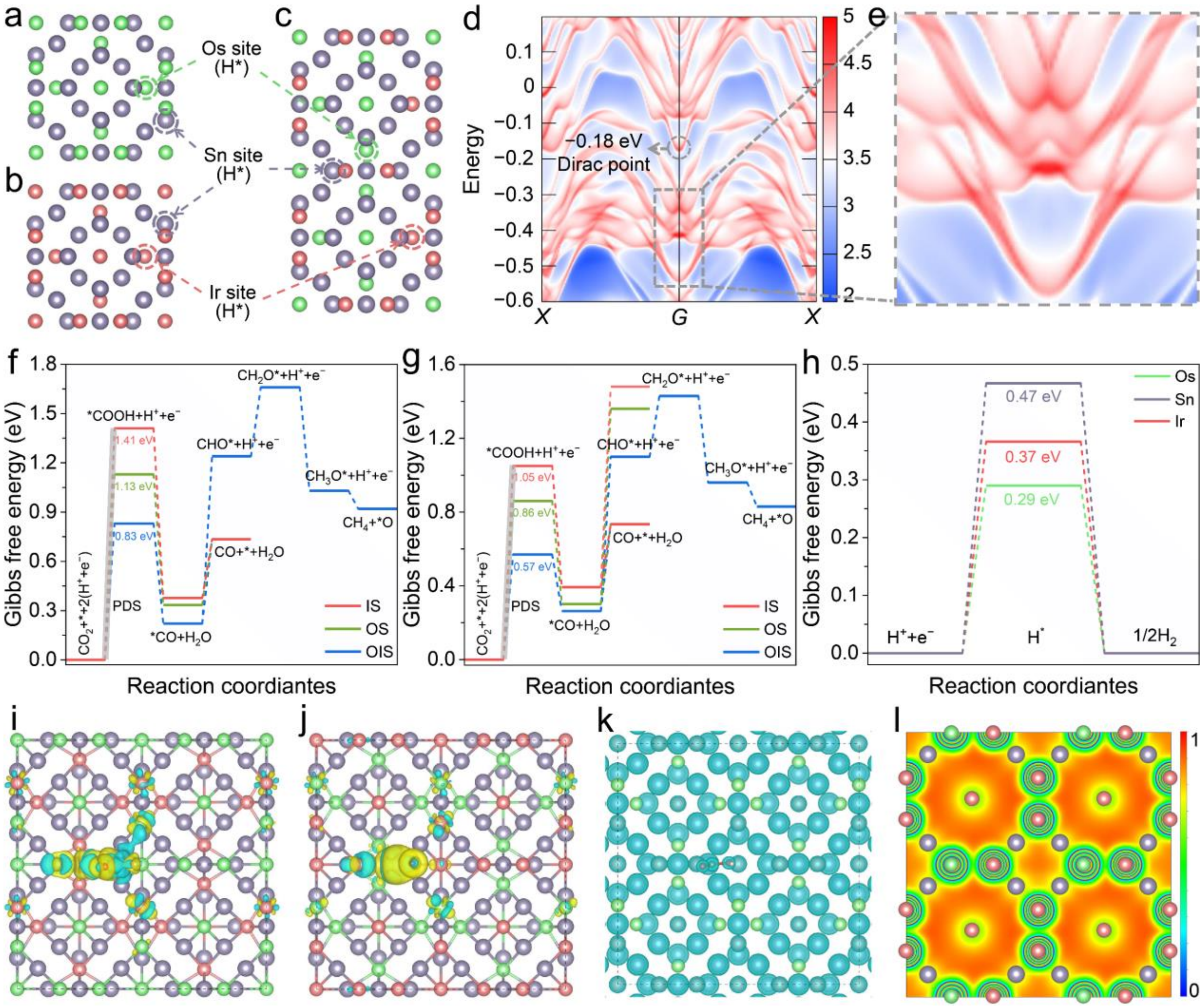

**Fig. 6.** The optimized atomic models of (a) OS, (b) IS, and (c) OIS in theoretical calculations. Relaxed adsorptive structure of *H on the (001) facet of OIS, namely the adsorption sites in (h). (d) DOS of the (001) facet of OIS after adsorption of $CO_2$ molecules. (e) Enlarged view of the TSS in d. The calculated Gibbs free energy diagrams for CRR on the (f) surface and (g) (001) facet of the OIS, OS, and IS. (h) Free energies of *H on the Sn, Os, and Ir sites of the OIS surface. 3D electron density difference distributions for adsorption of (i) *COOH and (j) *CO on the OIS (Cyan and yellow regions represent electron accumulation and loss.). (k) The electrostatic potential and (l) electron localization for adsorption of *COOH on the OIS surface (The gray, red, and green spheres represent Sn, Ir, and Os atoms.).

## 4. Conclusion

To sum up, we construct OIS with dual active sites through a facile solid-phase method. Compared with OS (25.50 μmol·g$^{-1}$·h$^{-1}$), the CO evolution activity of OIS is more than 5-fold improvement, reaching 119.3 μmol·g$^{-1}$·h$^{-1}$. Detailed surface energy bands together with in-situ characterizations reveal that the nontrivial topological surface states appreciably facilitate charge-separation/electron-enrichment and adsorption/activation of $CO_2$ molecules, rendering highly efficient reaction channels to promote the formation of *COOH and *CO, and subsequently desorption of *CO towards CO generation. Moreover, theoretical calculations indicate that the synergistic effect of Os−Ir dual-sites not only reduces the reaction barrier for the formation of *COOH and weakens the CO adsorption, but also retards the undesired HER, realizing high $IQE_{\mathrm{cr}}$ for CO. This work emphasizes the importance of tailoring nontrivial topological surface states for the rational design of high-activity photocatalysts.

**Table 1.** Carrier transport characteristics of OIS, OS, and IS semimetals at 300 K.

| | Materials | | | Slope | Hall coefficient | Carrier concentration | Mobility | Diffusion |
|---|---|---|---|---|---|---|---|---|
| | Length (m) | Width (m) | Height (m) | (Ω/T) | ($m^3$·$C^{-1}$) | ($m^{-3}$) | ($m^2$·$V^{-1}$·$s^{-1}$) | length (μm) |
| OIS | $4.08\times10^{-3}$ | $2.42\times10^{-3}$ | $1.25\times10^{-3}$ | $5.35\times10^{-7}$ | $6.69\times10^{-10}$ | $9.33\times10^{27}$ | $2.36\times10^{-4}$ | 0.082~0.26 |
| OS | $4.11\times10^{-3}$ | $2.54\times10^{-3}$ | $7.9\times10^{-4}$ | $9.88\times10^{-7}$ | $7.81\times10^{-10}$ | $7.99\times10^{27}$ | $1.12\times10^{-4}$ | 0.058~0.18 |
| IS | $3.35\times10^{-3}$ | $3.18\times10^{-3}$ | $1.3\times10^{-3}$ | $3.35\times10^{-5}$ | $4.36\times10^{-8}$ | $1.43\times10^{26}$ | $9.4\times10^{-5}$ | 0.056~0.13 |

**Acknowledgments**

This work is funded by the Natural Science Foundation of China (No. 11922415, 12274471), Guangdong Basic and Applied Basic Research Foundation (No. 2022A1515011168), Guangzhou Science and Technology Programmer (No. 2024A04J6415) and the State Key Laboratory of Optoelectronic Materials and Technologies (Sun Yat-Sen University, No. OEMT-2024-ZRC-02). The experiments and calculations reported were conducted at the Guangdong Provincial Key Laboratory of Magnetoelectric Physics and Devices, No. 2022B1212010008.

# Supporting Information

## Revealing the nontrivial topological surface states of catalysts for effective photochemical carbon dioxide conversion

*Kangwang Wang*[a], *Longfu Li*[a], *Peifeng Yu*[a], *Nannan Tang*[b], *Lingyong Zeng*[a], *Kuan Li*[a], *Chao Zhang*[a], *Rui Chen*[a], *Zaichen Xiang*[a], *Huichao Wang*[b], *Yongqing Cai*[c,*], *Kai Yan*[d,*], *Huixia Luo*[a,*]

[a] School of Materials Science and Engineering, State Key Laboratory of Optoelectronic Materials and Technologies, Key Lab of Polymer Composite & Functional Materials, Guangzhou Key Laboratory of Flexible Electronic Materials and Wearable Devices, Sun Yat-sen University, Guangzhou, 510275, China

[b] Guangdong Provincial Key Laboratory of Magnetoelectric Physics and Devices, School of Physics, Sun Yat-sen University, Guangzhou, 510275, China

[c] Joint Key Laboratory of the Ministry of Education, Institute of Applied Physics and Materials Engineering, University of Macau, Macau, Taipa, 999078, China

[d] School of Environmental Science and Engineering, Sun Yat-sen University, Guangzhou, 510275, China

E-mail: yongqingcai@um.edu.mo (Y. Cai), yank9@mail.sysu.edu.cn (K. Yan), luohx7@mail.sysu.edu.cn (H. Luo)

## Section I: Characterization

### ICP-OES measurements

Inductively coupled plasma optical emission spectrometer (ICP-OES) was tested on an Agilent 7500cx instrument with an attached laser ablation system.

### Hall effect measurements

The Hall measurements were performed in a physical property measurement system (PPMS). The Lorentz force is in equilibrium with the electrostatic force:

$$Bqv = Eq \tag{1}$$

$$Bv = E \tag{2}$$

$$U = Ed = Bvd \tag{3}$$

$$I = \frac{\mathrm{d}q}{\mathrm{d}t} = \frac{n\left(v\mathrm{d}t\right)hdq}{\mathrm{d}t} = nvhdq \tag{4}$$

$$U = Bd\cdot\frac{I}{ndhq} = \frac{1}{nq}\cdot\frac{IB}{h} \tag{5}$$

$$R = \frac{1}{nq}\cdot\frac{B}{h} = \frac{1}{nqh}\cdot B \tag{6}$$

The Hall data measured by PPMS is in *R*-*B* curve and is linearly fitted with slope $m = \frac{1}{nqh}$.

Hall coefficient ($R_H$): $$R_H = m\cdot h \tag{7}$$

Carrier concentration ($n$): $$n = \frac{1}{R_H q} \tag{8}$$

where $q$ is the electric charge of electrons, $h$ dotes as the height of the sample, $B$ refers to the magnetic field strength, and $R$ represents the resistance.

### Carrier diffusion lengths

The carrier diffusion lengths ($L_D$) were estimated based on the equation:

$$L_D = \left(\frac{k_B T}{e}\times\mu\times\tau\right)^{\frac{1}{2}} \tag{9}$$

where $k_B$ is the Boltzmann's constant, $T$ is the absolute temperature, $\mu$ is the carrier mobility, and $\tau$ is the carrier lifetime [1, 2].

**X-ray Absorption Spectra (XAS) measurements**

The Ir $L_3$-edge and Os $L_3$-edge XAS measurements were conducted at the BL14W1 station within the Shanghai Synchrotron Radiation Facility (SSRF), which operates at 3.5G eV with a maximum current of 250 mA. EXAFS fitting was applied through Athena and Artemis software. WT spectra was also employed using the software package developed by Funke and Chukalina using Morlet wavelet with $\kappa$ = 20 and $\sigma$ = 1. The data in XANES region of the absorption coefficient were examined via applying the same procedure for pre-edge line fitting, post edge curve fitting, and edge step normalization to all data.

**Apparent quantum efficiency (*AQE*) measurements**

The *AQE* was determined under the above photocatalytic reaction conditions except that Xe lamp was used as the irradiation source. The wavelength-dependent *AQE* was measured under the same photocatalytic reaction condition, except for the monochromatic light wavelengths (380, 400, 420, 440, 460, 480, 500, and 520 nm). At the same time, the *AQE* and corresponding *IQE* ($IQE_{cr}$) were roughly calculated from the equation:

$$AQE = \frac{N_{\text{electrons}}}{N_{\text{photons}}} \times 100\% \qquad \text{ES1}$$

$$N_{\text{electrons}} = n_{\text{electrons}} \times N_{\text{A}}$$

where $N_A$ is Avogadro constant ($6.02 \times 10^{23}$ mol$^{-1}$).

$$N_{\text{photons}} = \frac{ItS_1}{Q}; Q = \frac{hc}{\lambda}; I = \frac{P}{S_2} \qquad \text{ES2}$$

where $I$, $S_1$, $t$, and $Q$ represent light intensity (W·m$^{-2}$), irradiation area (in this work, $15.89 \times 10^{-4}$ m$^2$), irradiation time (s), and photon energy (J), respectively. $P$ and $S_2$ are the light power (W) and aperture area (in this work, $3.14 \times 10^{-4}$ m$^2$) of the light power meter, respectively [1,2].

$$N_{\text{photons}} = \frac{\frac{P}{S_2} t S_1}{\frac{hc}{\lambda}} = \frac{P t S_1 \lambda}{S_2 hc} = \frac{P S_1}{S_2} \frac{\lambda t}{hc} = E \frac{\lambda t}{hc} \qquad \text{ES3}$$

where $h$, $c$, and $\lambda$ represent Planck's constant ($6.626 \times 10^{-34}$ j·s), light speed ($3 \times 10^{-34}$ m·s$^{-1}$), and monochromatic light wavelength, respectively.

$$IQE_{\text{cr}} = \frac{AQE}{A} \times 100\% \qquad \text{ES4}$$

where A is the absorption of sample [3].

The values of $E$ at different wavelengths ($\lambda$ = 380, 400, 420, 440, 460, 480, 500, and 520 nm) was calculated to be 0.73, 0.69, 0.67, 0.73, 0.81, 0.62, 0.67, and 0.71 W by Xe lamp (300 W Xe lamp, Beijing Perfect light Technology Co. Ltd, PLS-SXE300D) source.

Taking OIS at $\lambda$ = 420 nm as an example:

$$AQE_{(CH_4)} = \frac{8 \times 0.215 \times 10^{-6} \times 6.02 \times 10^{23}}{5.12 \times 10^{21}} \times 100\% = 0.02\%$$

$$AQE_{(CO)} = \frac{2 \times 5.967 \times 10^{-6} \times 6.02 \times 10^{23}}{5.12 \times 10^{21}} \times 100\% = 0.14\%$$

$$\text{Total } AQE = AQE_{(CH_4)} + AQE_{(CO)} = 0.16\%$$

$$IQE_{\text{cr}} = \frac{\text{Total } AQE}{A} = \frac{0.16\%}{0.826} = 0.19\%$$

**In-situ DRIFTS measurements**

In-situ DRIFTS measurements were performed on a Thermo Scientific Nicolet iS50 system equipped with a liquid nitrogen-cooled mercury-cadmium-telluride (MCT) detector. Each spectrum was acquired in the range of 2000 ~ 500 cm$^{-1}$ with a resolution of 5 cm$^{-1}$. The samples were compressed at the bottom of an in-situ attenuated total reflectance (ATR) cell and then degassed for 1 h in an Ar atmosphere at 120 °C to remove absorbed material before photocatalytic experiments were performed. After cooling to room temperature, humidified $CO_2$ ($CO_2$:$H_2O$ = 8:1) was fed into the specimen chamber to saturate the photocatalyst surface with adsorption. A 300 W Xe lamp was used as the light source. In-situ DRIFTS spectra were collected after certain irradiation times (e.g., 0, 10, 20, 30, 40, 50, and 60 min).

**In-situ NAP-XPS measurements**

In-situ NAP-XPS measurements were obtained on the beamline BL02B1 of SSRF equipped with a 300 W Xe lamp as the illumination source.

**Computational method of charge distribution**

All of the calculations are performed in the framework of the spin-polarized DFT with the projector augmented plane-wave method, as implemented in the Vienna ab initio simulation package (VASP) and Materials Studio 2019. The projector augmented wave (PAW) method was adopted to describe the interactions between the electrons and ions. The Perdew-Burke-Ernzerhof function (PBE) within the generalized gradient approximation (GGA) was applied to describe the exchange-correlation interaction. The cutoff energy of the plane-wave basis set was 570 eV for expanding the electron wave function. The energy criterion was set to $10^{-5}$ eV in iterative solution of the Kohn-Sham equation. The Brillouin-zone sampling was conducted using Monkhorst-Pack (MP) grids of special points with the separation of 0.04 $Å^{-1}$. All the structures were relaxed until the residual forces on the atoms have declined to less than 0.02 eV $Å^{-1}$. The $\Delta G$ was defined as:

$$\Delta G = \Delta E + \Delta E_{\mathrm{ZPE}} - T\Delta S \tag{10}$$

where $\Delta E$ is the energy difference between the reactants and product obtained through DFT calculations. $\Delta E_{\mathrm{ZPE}}$ and $\Delta S$ are the changes in the zero-point energies (ZPE) and entropy. $T$ represents the temperature and was set as 298.15 K.

## Section II: Figures and Tables

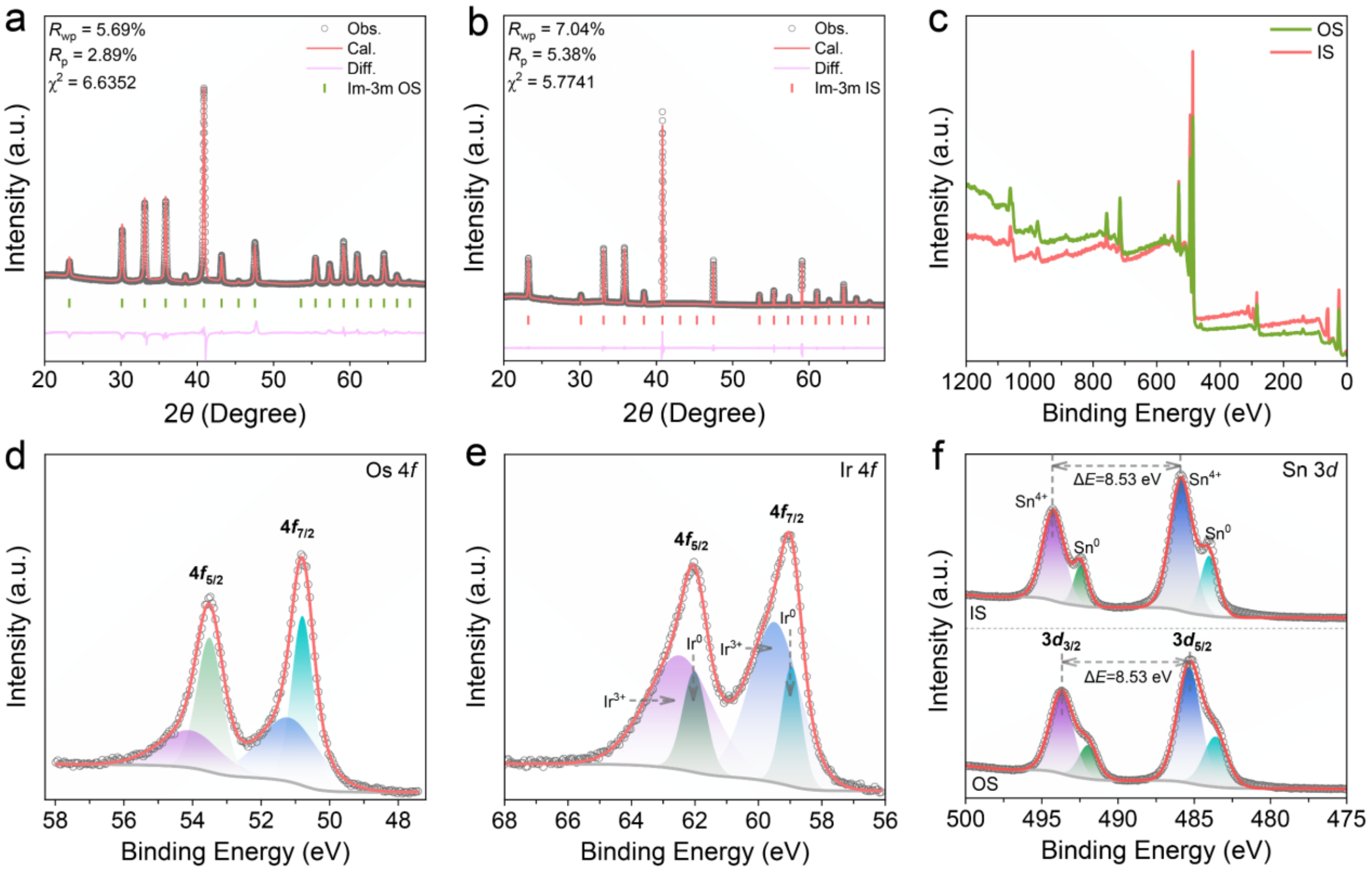

**Fig. S1.** Rietveld refinement of XRD data for (a) OS and (b) IS, respectively. (c) XPS full-scan survey, (d) Os 4*f*, (e) Ir 4*f*, and (f) Sn 3*d* spectra of IS and OS, respectively.

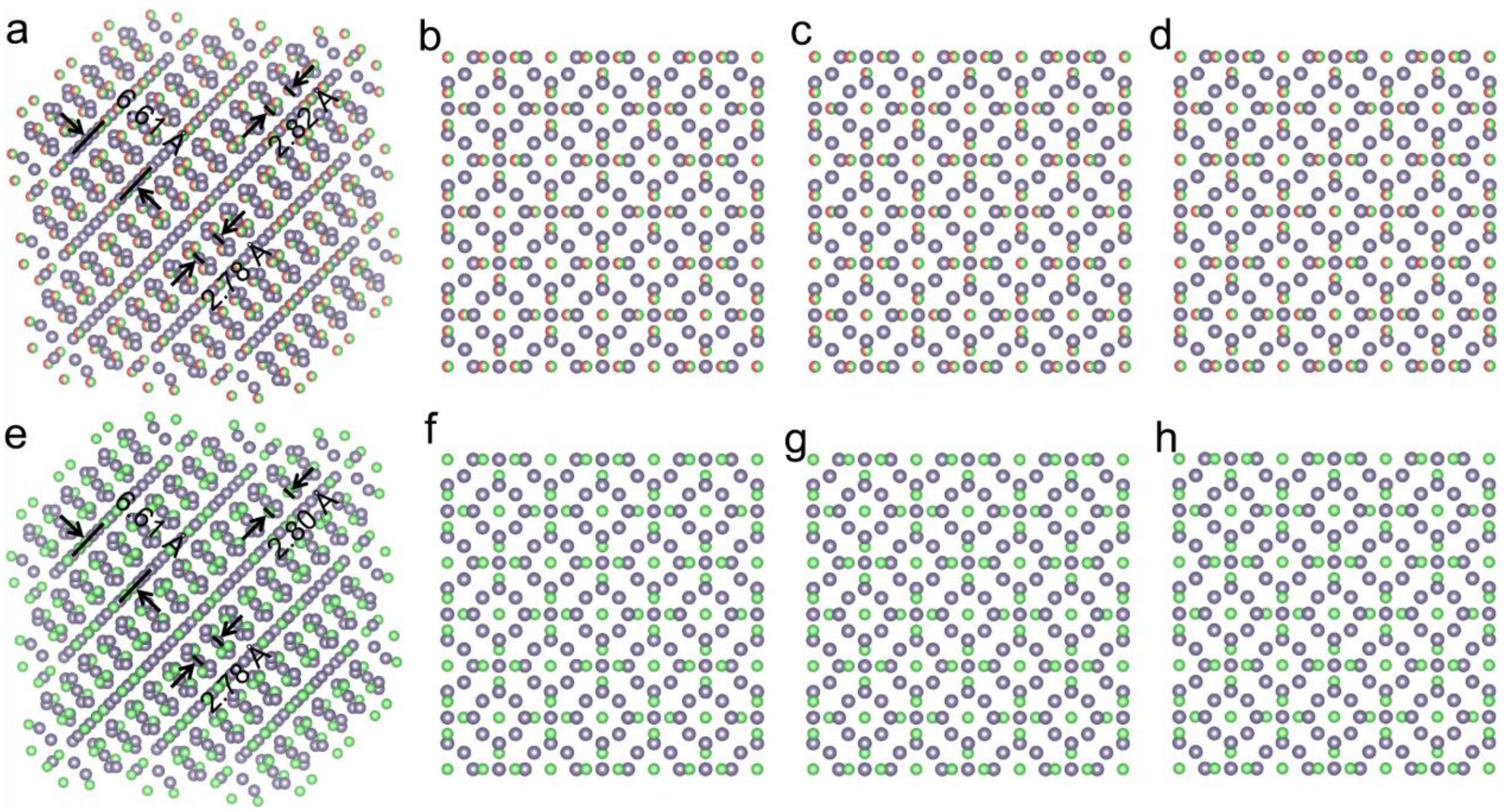

**Fig. S2.** Crystal structures of OIS, viewed from (a) [3-73], (b) [100], (c) [001], and (d) [010] directions, respectively. OIS in a cubic structure (space group Im-3m) with lattice parameter values a = b = c = 9.36816 Å. It could be seen that OS and IS, OIS only differ by the substitution

of Os and Ir atoms of in OIS. The gray, red, and green spheres represent Sn, Ir, and Os atoms, respectively. Crystal structures of OS, viewed from (e) [3-73], (f) [100], (g) [001], and (h) [010] directions, respectively. OS in a cubic structure (space group Im-3m) with lattice parameter values a = b = c = 9.43818 Å.

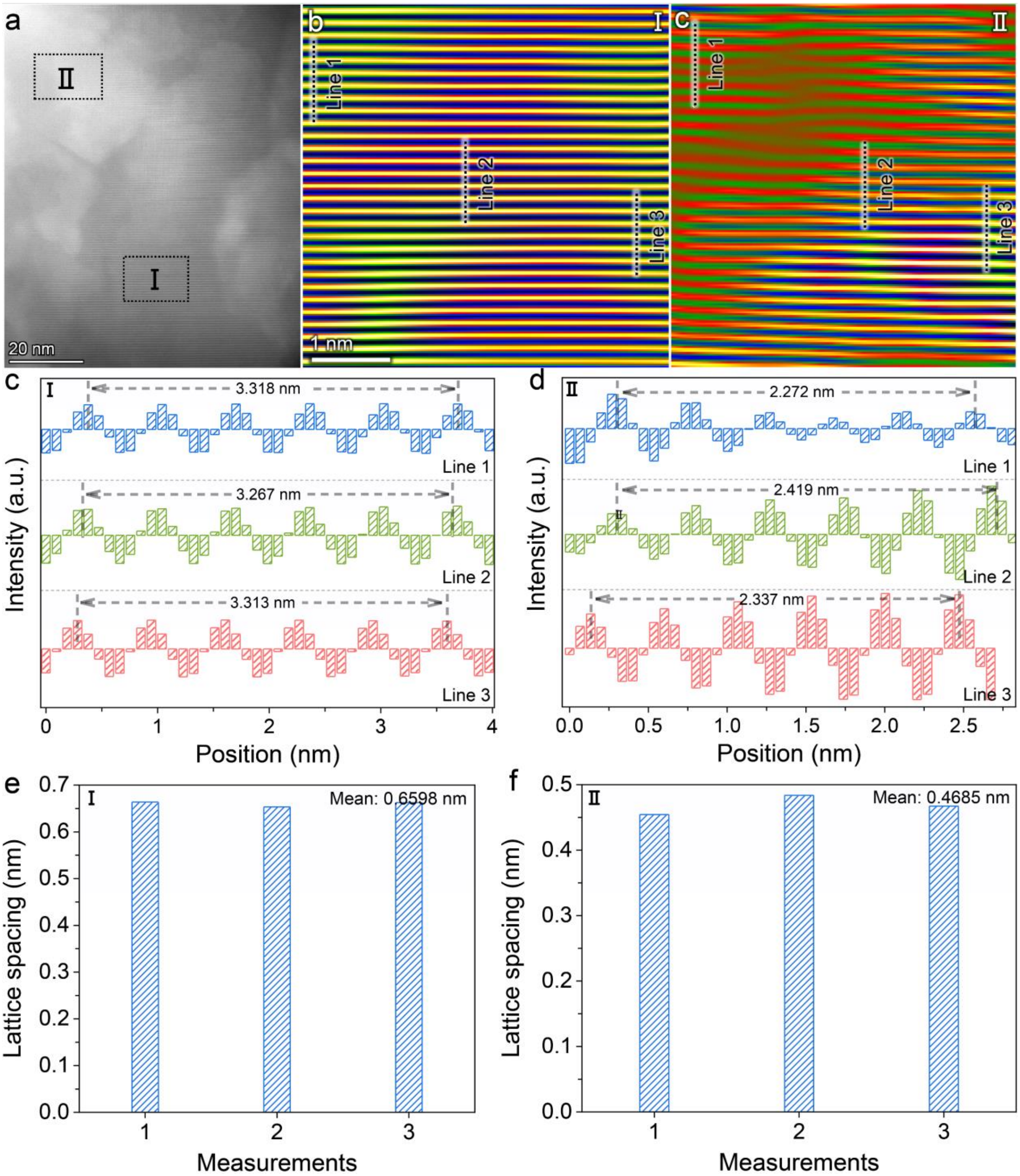

**Fig. S3.** Top panels: lattice fringes obtained by applying FFT filters at different directions to the STEM image of OIS [3-73] facet in the main text. Middle panels: the brightness along the three lines marked in the top panels, along with the measured distance across several periods.

Below panels: the calculated lattice spacings according to the measurements in the middle panels.

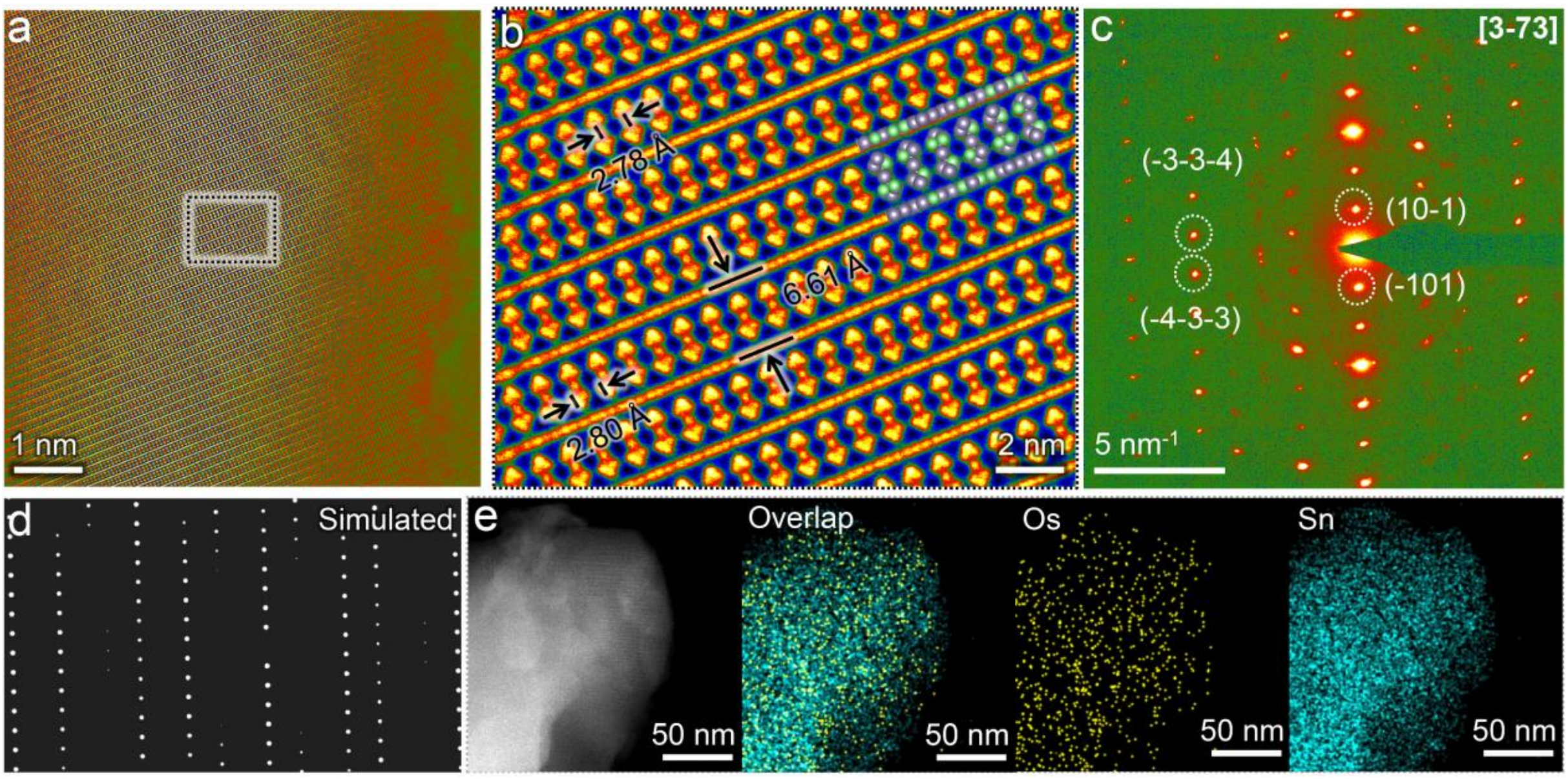

**Fig. S4.** (a) HAADF-STEM image taken in the [3-73] direction of OS. (b) Magnified HAADF-STEM image taken from the corresponding area in (a) (top) of OS from [3-73] direction (bottom). (c) SAED pattern of OS along the [3-73] direction and (d) Simulated electron diffraction pattern corresponding to the (c) panel. (e) EDS mapping of OS.

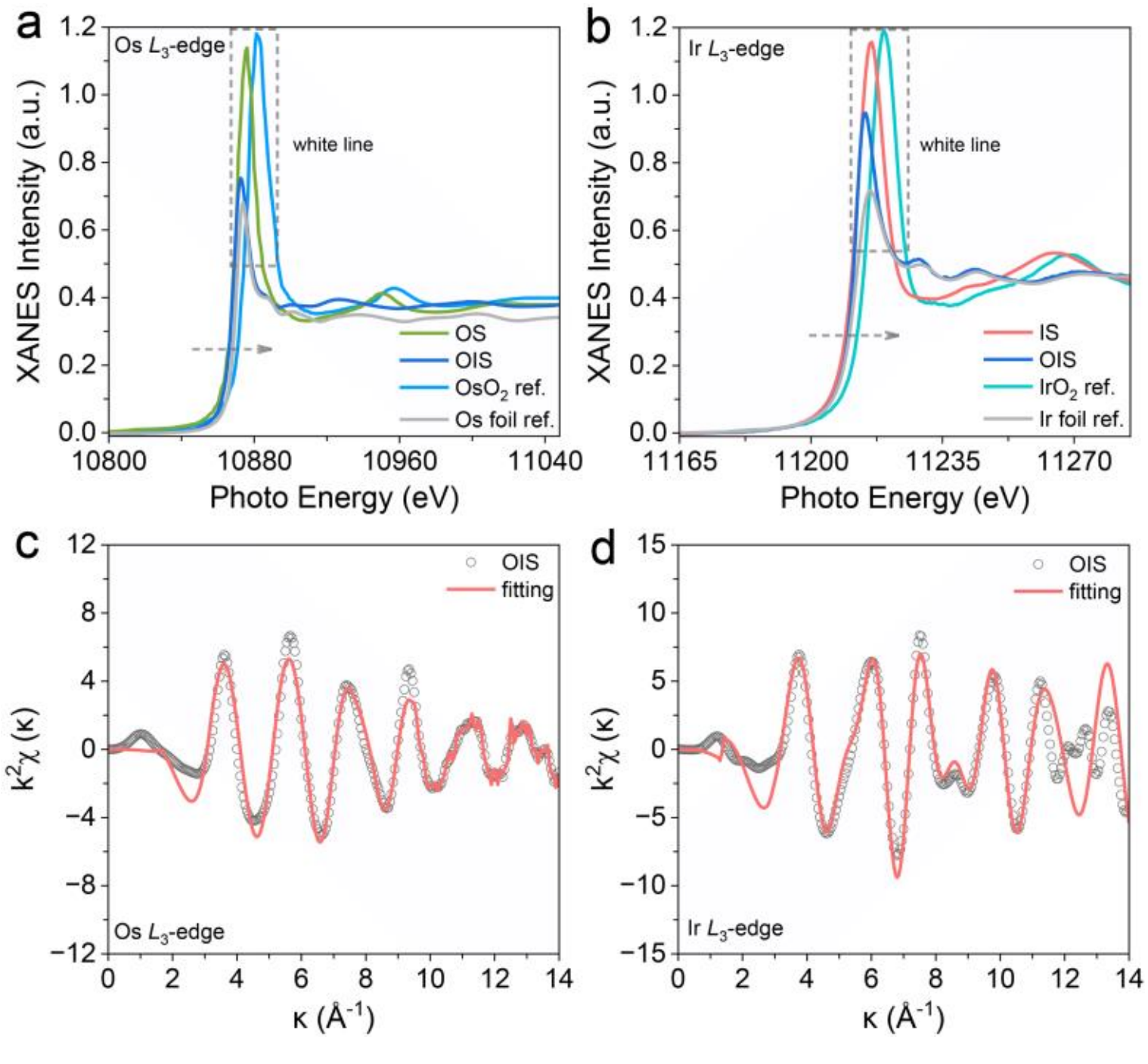

**Fig. S5.** (a) Os and (b) Ir $L_3$-edge XANES spectra of OIS, OS, and IS. Fitting results of (c) Os and (d) Ir $L_3$-edge EXAFS spectra in *k*-space for OIS.

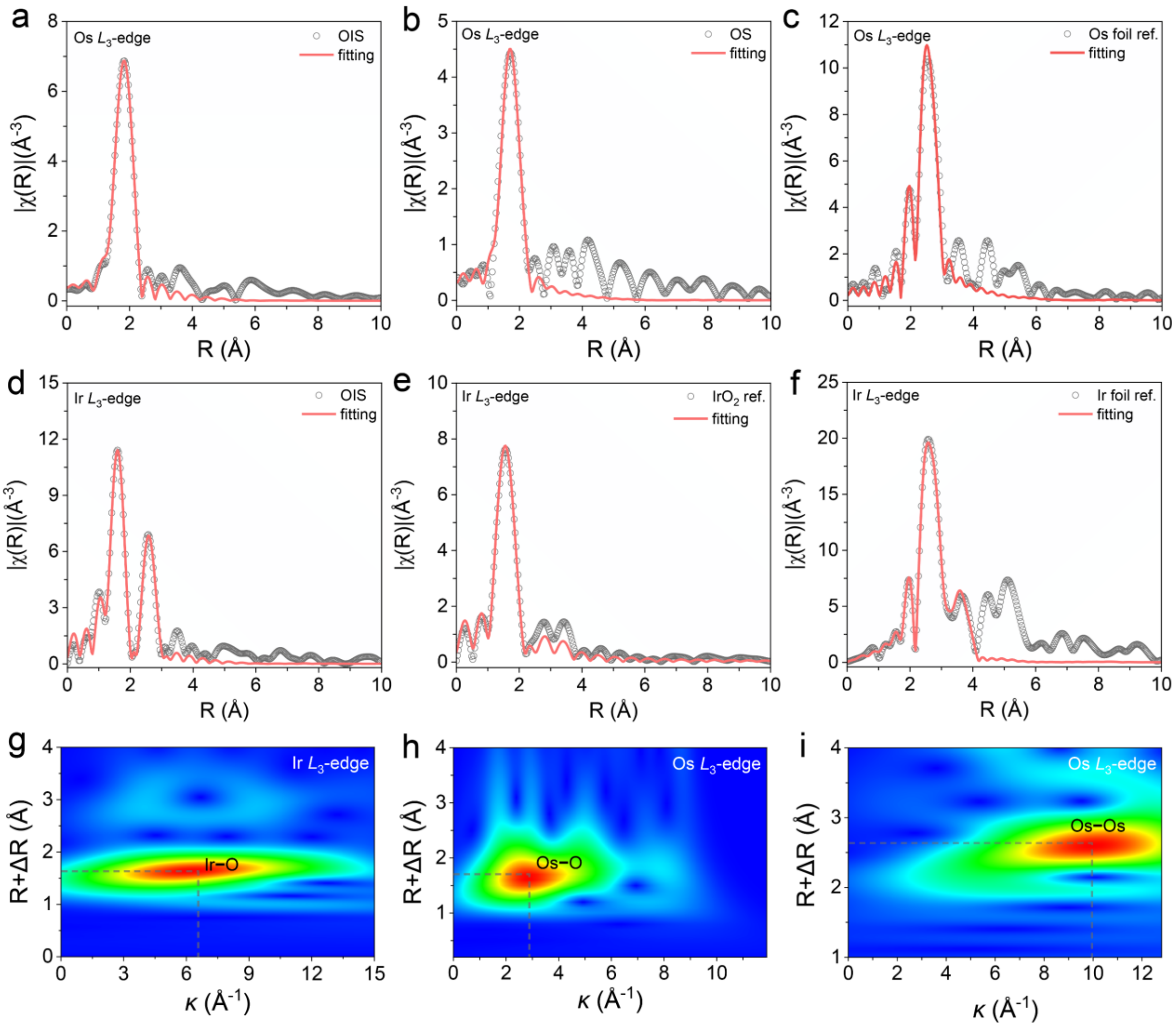

**Fig. S6.** Fitting results of EXAFS spectra in *R*-space for (a) OIS, (b) OS, and (c) Os foil ref.. Fitting results of EXAFS spectra in *R*-space for (d) OIS, (e) $IrO_2$ ref., and (f) Ir foil ref.. WT for (k) Ir $L_3$-edge of $IrO_2$ ref. and Ir $L_3$-edge of (h) $OsO_2$ ref. and (i) Os foil ref..

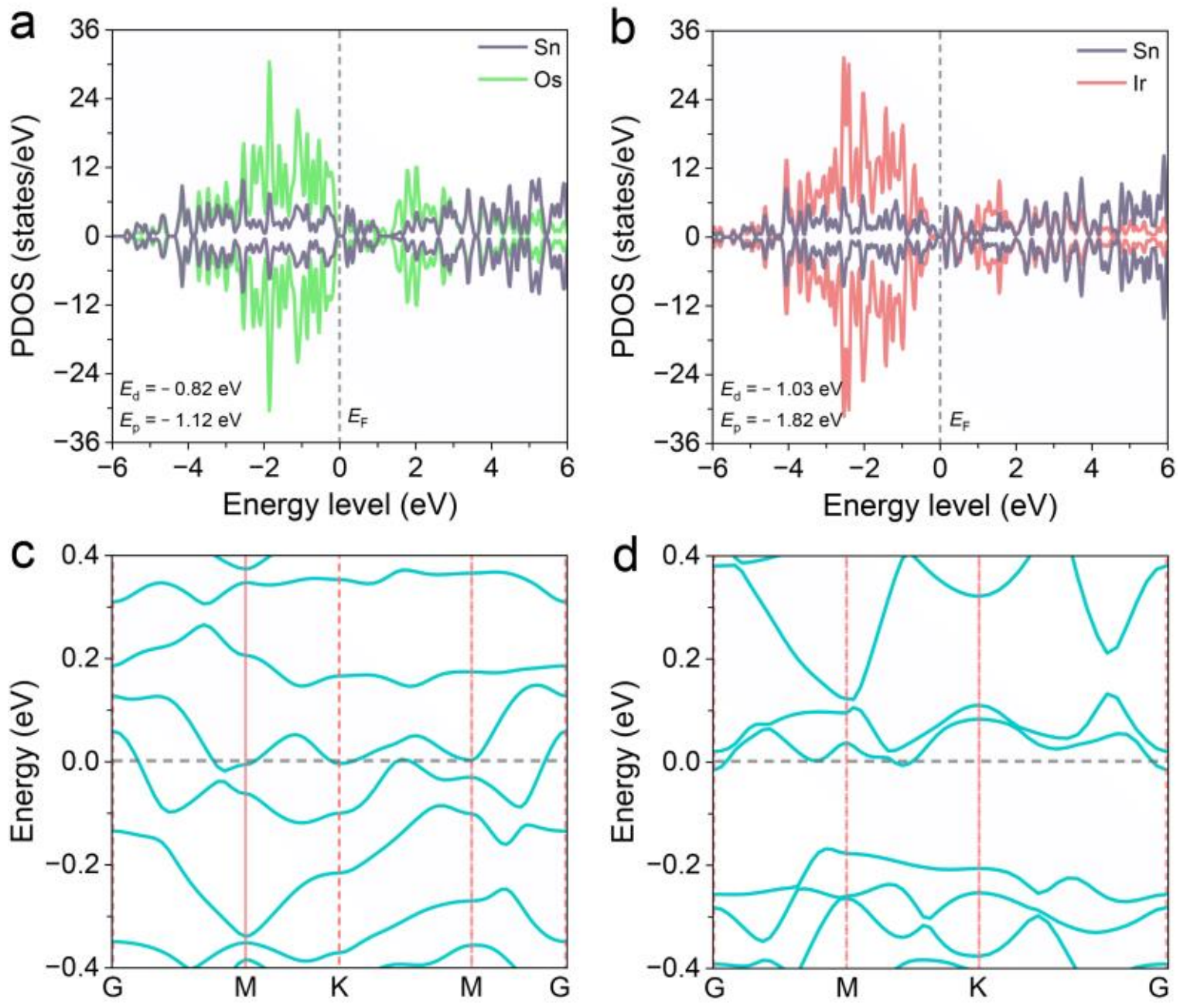

**Fig. S7.** The calculated DOS (a) OS and (b) IS. The surface energy band of the (c) (001) and (d) (110) facet of OIS.

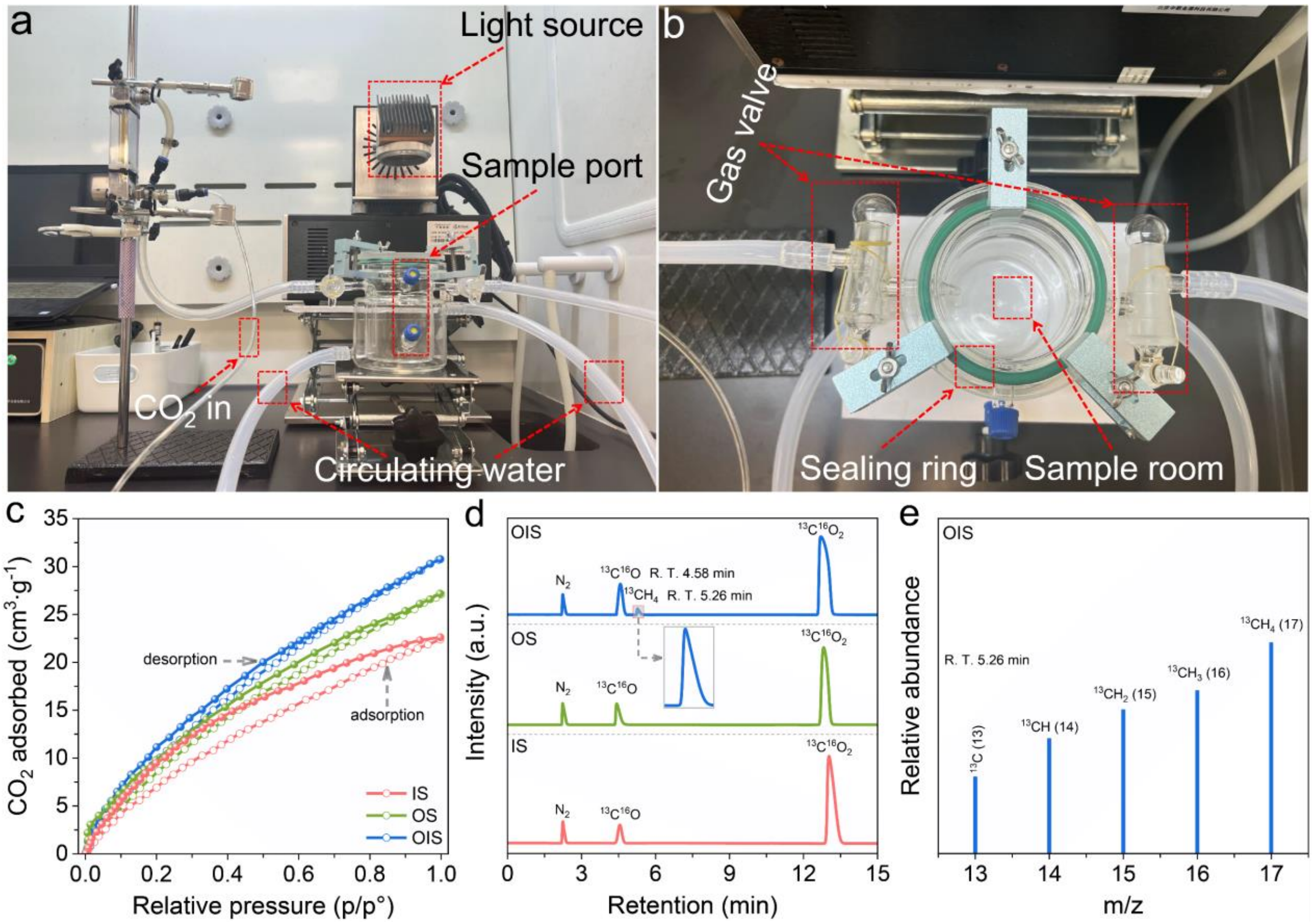

**Fig. S8.** (a,b) The photos for the reaction device of photocatalytic CRR test. (c) $CO_2$ adsorption isotherms of OIS, OS, and IS at 298 K. (d) GC spectra of OIS, OS, and IS from the photocatalytic system. (e) GC-MS spectra of $^{13}C$, $^{13}CH$, $^{13}CH_2$, $^{13}CH_3$, and $^{13}CH_4$ produced from the photocatalytic reduction of $^{13}CO_2$ for different photocatalytic system.

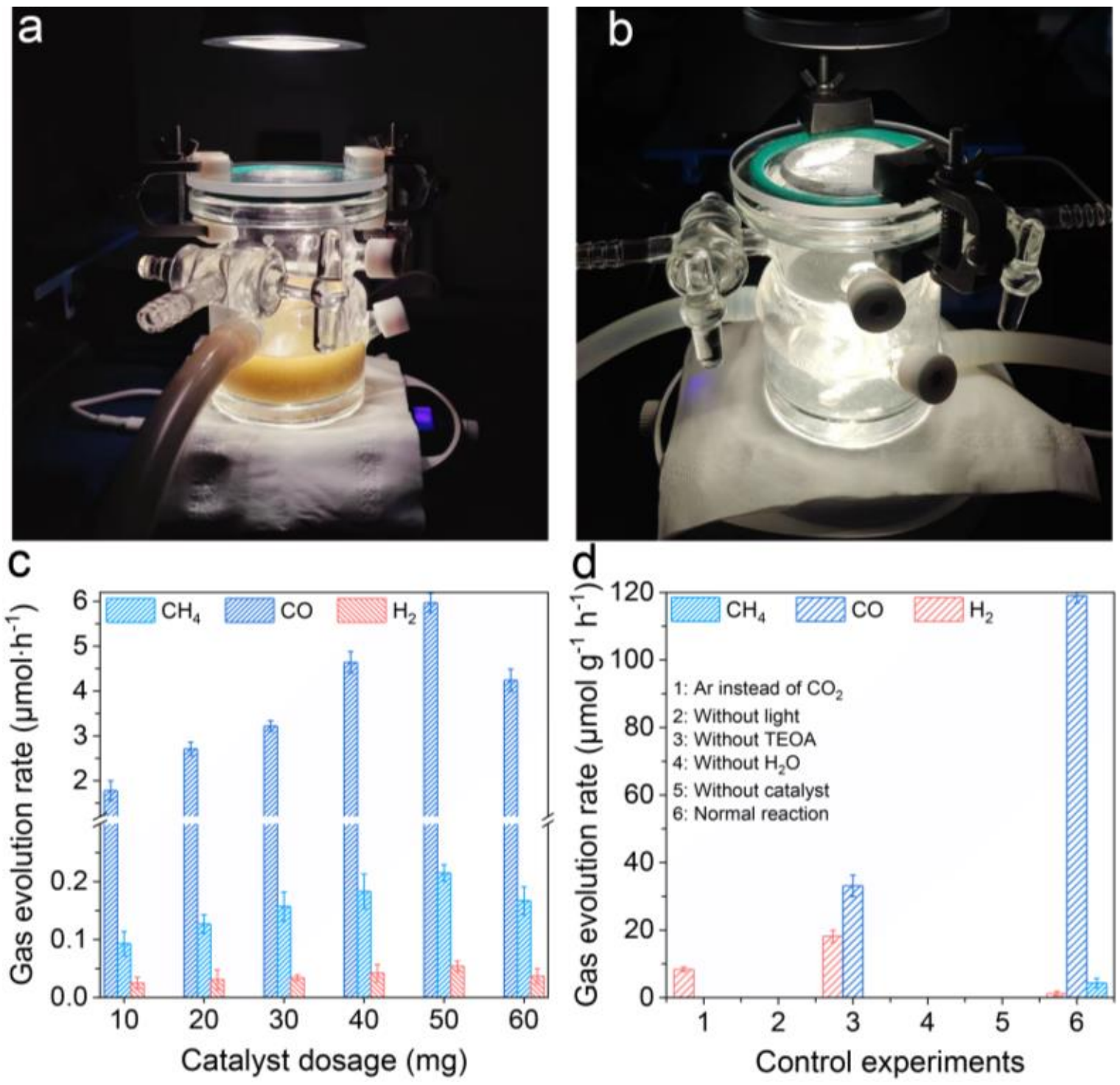

**Fig. S9.** Digital images of the photocatalytic CRR to CO measurements under Xe lamp using OIS and IS as the catalyst. (c) Mass-loading for $CO_2$ photoreduction. (d) Control experiments in several conditions.

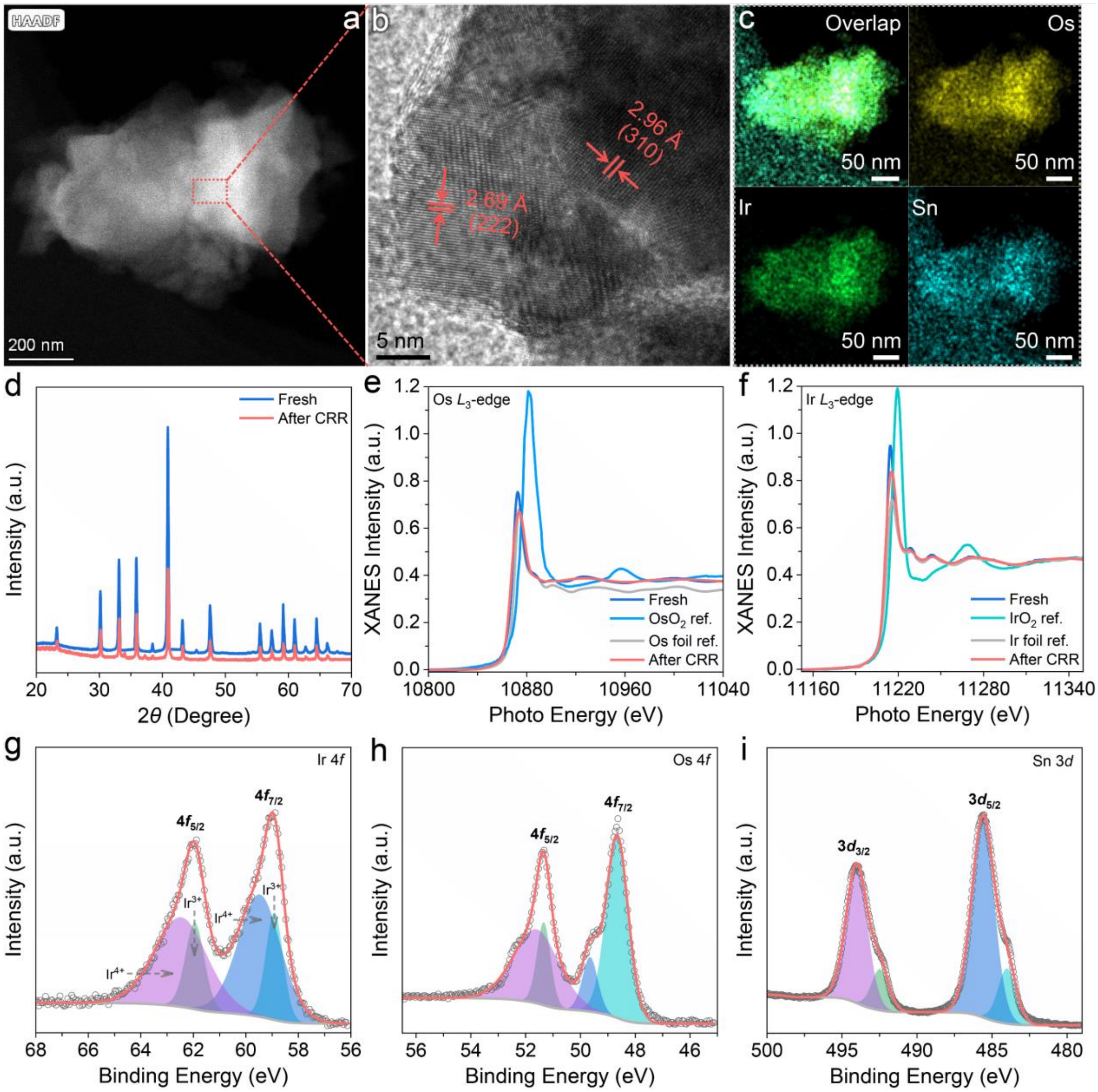

**Fig. S10.** (a,b) TEM and (c) EDS mapping images of OIS after photocatalytic CRR (nine cycles). (d) XRD pattern, (e) Os and (f) Ir $L_3$-edge XANES spectra, (g) Ir 4*f*, (h) Os 4*f*, and (i) Sn 3*d* XPS spectra of OIS after CRR (nine cycles).

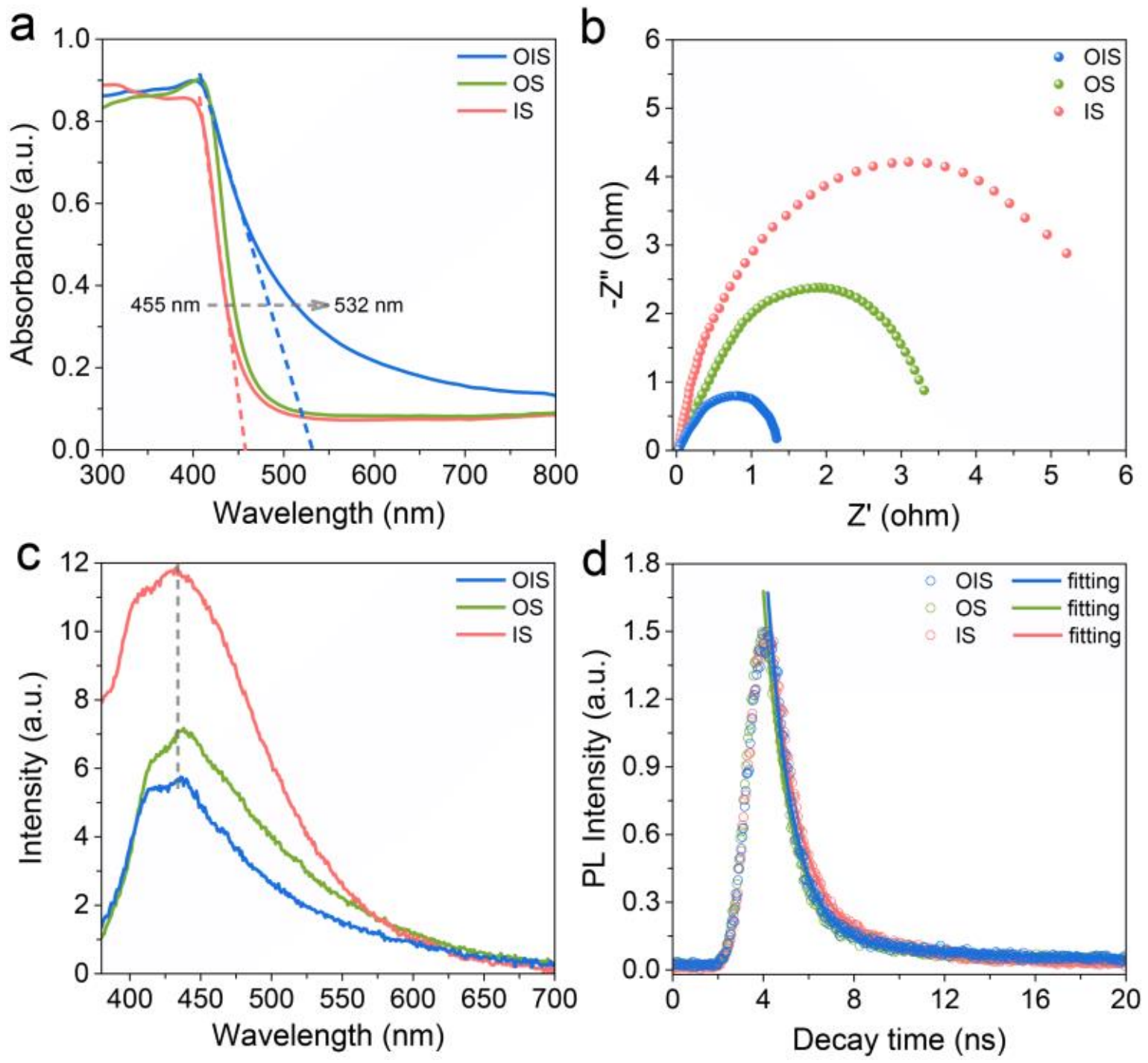

**Fig. S11.** (a) UV-vis DRS, (b) EIS, (c) PL ($\lambda_{ex}$ = 320 nm), and (d) TRPL spectra of OIS, OS, and IS.

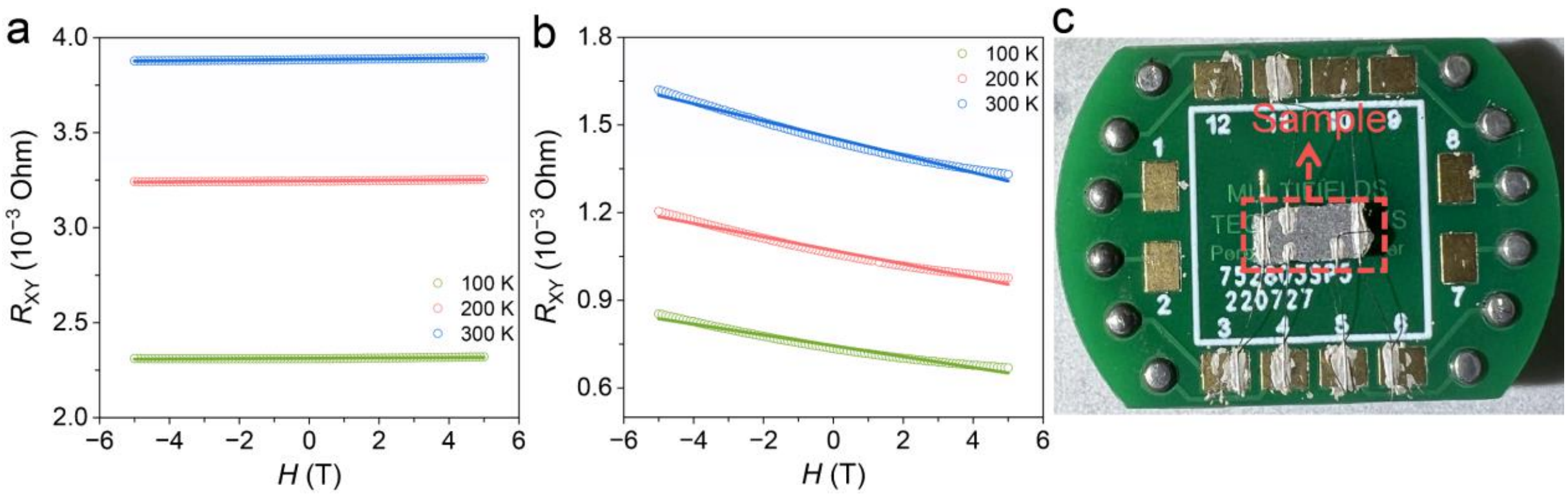

**Fig. S12.** Hall resistivity versus magnetic field of (a) OS and (b) IS at different temperature. (c) Schematic diagram of electronic components for Hall resistivity test of OIS.

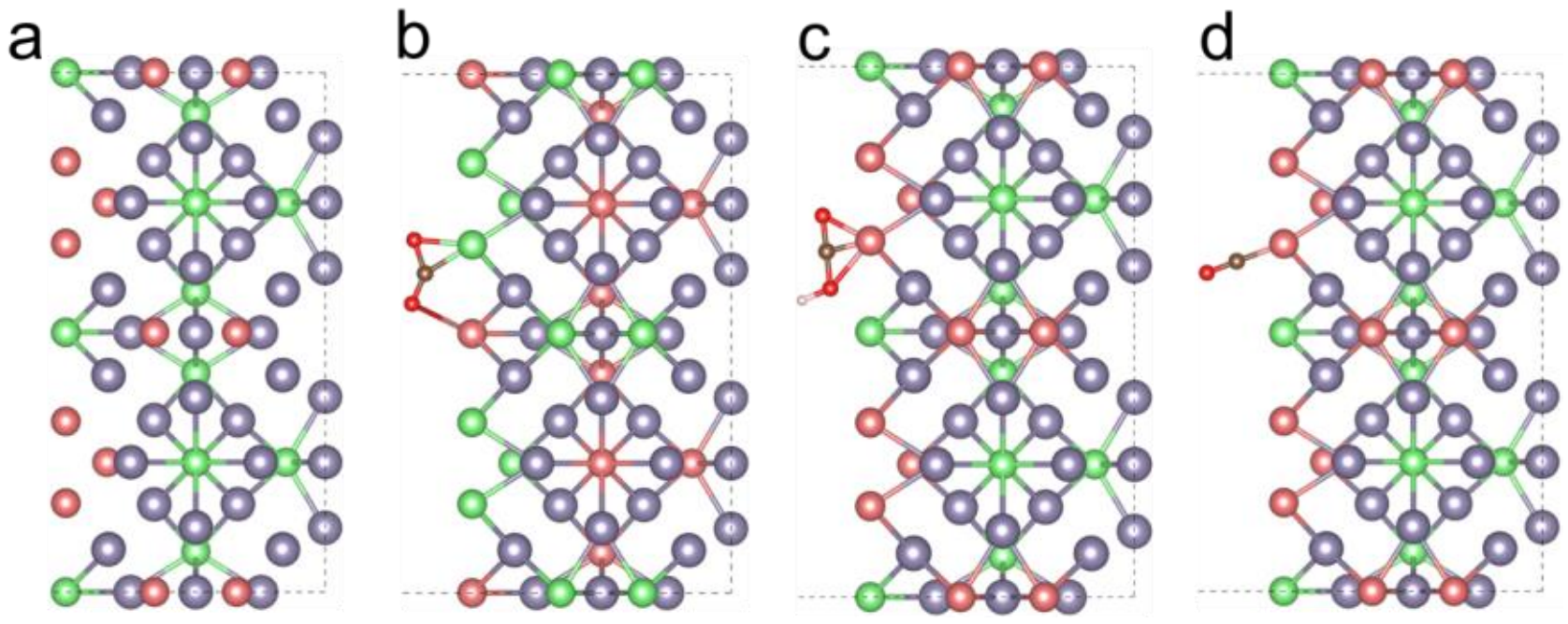

**Fig. S13.** Schematic reaction CRR to CO pathway projected on the OIS surface (side view).

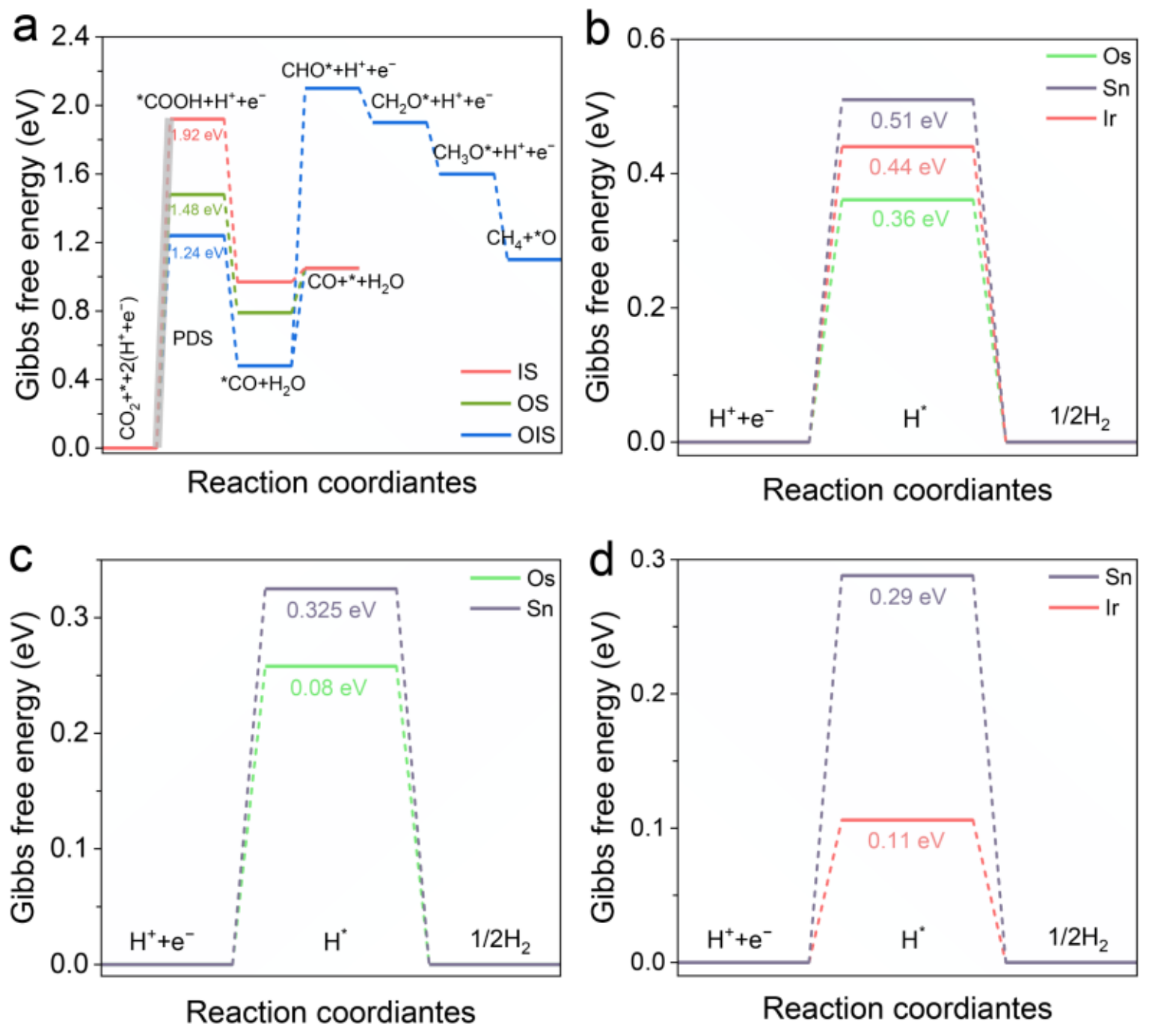

**Fig. S14.** (a) The calculated Gibbs free energy diagrams for CRR on the (110) facet of OIS, OS, and IS. (b) Free energies of *H on Sn, Os, and Ir sites of OIS (001) facet. (c) Free energies of *H on Os and Sn sites of OIS surface. (d) Free energies of *H on Ir and Sn sites of OIS surface.

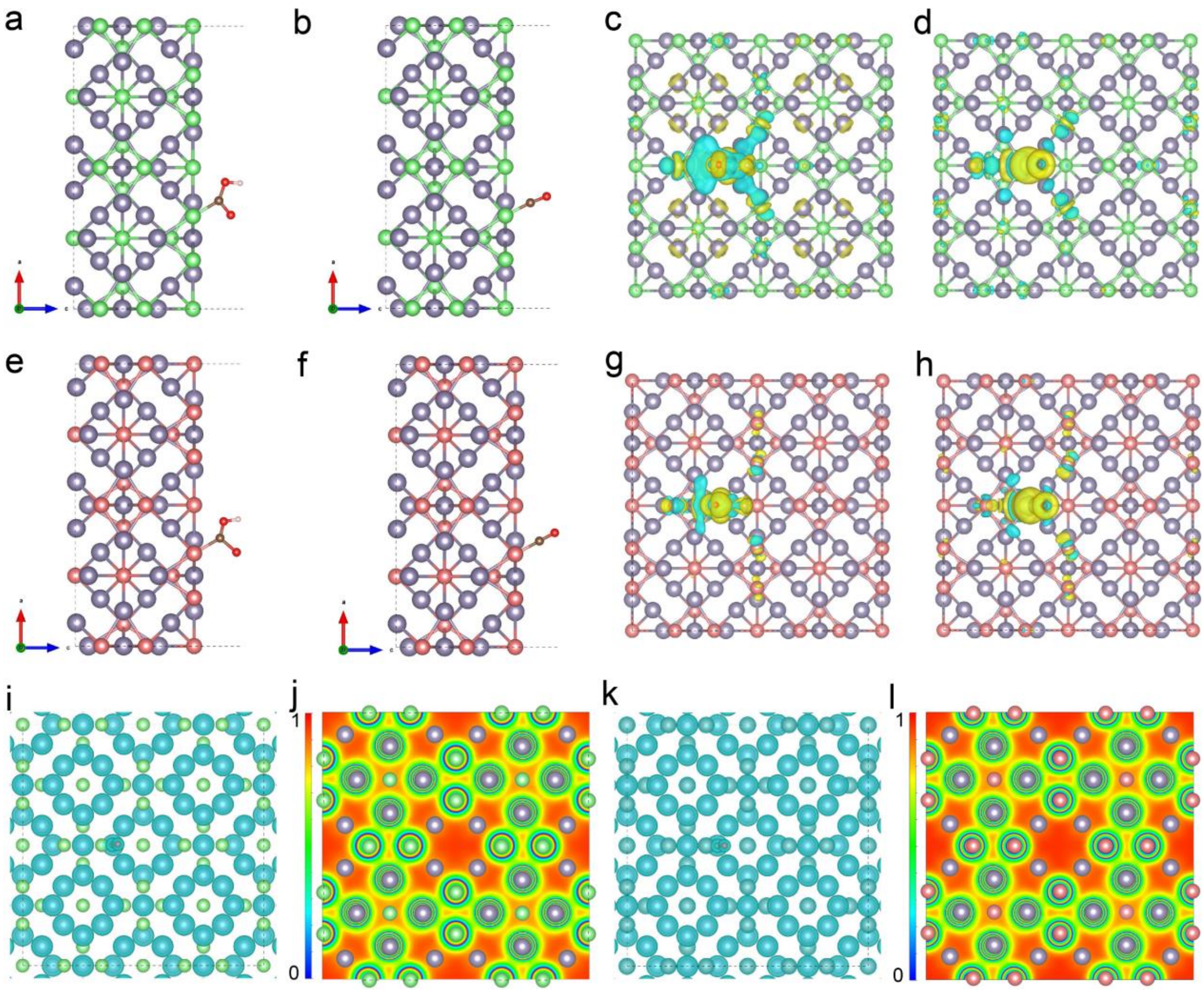

**Fig. S15.** Schematic reaction CRR to CO pathway projected on the (a,b) OS and (e,f) IS surface. Charge density differences of (c) *COOH and (d) *CO on the OS. Charge density differences of (g) *COOH and (h) *CO on the IS (Cyan and yellow regions represent electron loss and accumulation). (i) The electrostatic potential and (j) electron localization for adsorption of *CO on the OS interface. (k) The electrostatic potential and (l) electron localization for adsorption of *CO on the IS interface.

**Table S1.** Os and Ir contents in OIS, OS, and IS based on XPS and ICP data.

| Catalysts | XPS (Atomic fraction (%)) | | | | ICP-OES (Mass fraction (%)) | | |
|---|---|---|---|---|---|---|---|
| | Os | | Ir | | Os | Ir | Sn |
| | $Os^{3+}$ | $Os^{0}$ | $Ir^{3+}$ | $Ir^{0}$ | | | |
| OIS | 34.76 | 65.24 | 35.65 | 64.35 | 24.4 | 24.8 | 50.8 |
| OS | 57.82 | 42.18 | / | / | 49 | / | 51 |
| IS | / | / | 44.38 | 55.62 | / | 48.3 | 51.7 |

Note: the data of OIS was normalized.

**Table S2.** A summary of photocatalytic CRR performances by the reported catalysts.

| Catalyst | Reaction medium | CO rate (μmol·g$^{-1}$·h$^{-1}$) | Sel$_{CO}$ (%) | Ref. |
|---|---|---|---|---|
| AuPd/(101)$TiO_2$ | $H_2O$+TEOA | 7.5 | 34.9 | [3] |
| $FeCoS_2$ | $H_2O$ | 4.13 | 17 | [4] |
| CT-COF | $H_2O$ | 102.7 | ~100 | [5] |
| CdS/$BiVO_4$ | $H_2O$ | 1.75 | 81.8 | [6] |
| TAPBBCOF | $H_2O$ | 24.6 | 95.6 | [7] |
| Co/BNF | $H_2O$ | 32 | 87.2 | [8] |
| Carbon doped $In_2S_3$ | $H_2O$ | 9 | 17.4 | [9] |
| $N_v$-rich-CN | $H_2O$ | 6.61 | 97 | [10] |
| CGS/GS | $H_2O$+TEOA | 51.67 | 12.6 | [11] |
| g-$C_3N_4$-N-$TiO_2$ | $H_2O$ | 1.23 | 79.3 | [12] |
| $AgInP_2S_6$ | $H_2O$ | 6.67 | 23.5 | [13] |
| CP5 | $H_2O$+TEOA | 47.37 | 98.3 | [14] |
| 12FLTC/BCN | $H_2O$ | 2.4 | 94.9 | [15] |
| CdS-Co/Au | $H_2O$ | 9.2 | 96.1 | [16] |
| $Cu^{\delta+}$/$CeO_2$-$TiO_2$ | $H_2O$ | 3.47 | 3.5 | [17] |
| Copper Oxide | $H_2O$+TEOA | 27.3 | 40.44 | [18] |
| O/BN | $H_2O$ | 12.5 | 15.8 | [19] |
| OIS | $H_2O$+TEOA+MCN | 119.3 | 99.1 | This work |
| OS | | 25.5 | 69.57 | |
| IS | | 8.65 | 35.75 | |
| OIS | $H_2O$+MCN | 33.1 | 64.5 | |

**Table S3.** The calculated AQE of OIS at different wavelengths.

| Wavelength (nm) | $N_{photons}$ | CO production (μmol) | $CH_4$ production (μmol) | Ave. $AQE_{CO}$ (%) | Ave. $AQE_{CH4}$ (%) | Ave. $AQE_{(Total\ C)}$ (%) | Absorption | $IQE_{cr}$ (%) |
|---|---|---|---|---|---|---|---|---|
| | | 6.54 | 0.36 | | | | | |
| 380 | $5.024\times10^{21}$ | 6.75 | 0.40 | 0.1573 | 0.0351 | 0.19 | 0.8820 | 0.2182 |
| | | 6.40 | 0.34 | | | | | |
| | | 6.34 | 0.28 | | | | | |
| 400 | $4.998\times10^{21}$ | 6.00 | 0.23 | 0.1520 | 0.0260 | 0.18 | 0.8960 | 0.1986 |
| | | 6.59 | 0.31 | | | | | |
| | | 5.97 | 0.17 | | | | | |
| 420 | $5.119\times10^{21}$ | 6.27 | 0.21 | 0.1403 | 0.0202 | 0.16 | 0.8260 | 0.1944 |
| | | 5.66 | 0.28 | | | | | |
| | | 5.47 | 0.22 | | | | | |
| 440 | $5.817\times10^{21}$ | 5.20 | 0.16 | 0.1084 | 0.0160 | 0.12 | 0.6580 | 0.1891 |
| | | 5.04 | 0.20 | | | | | |
| | | 4.83 | 0.19 | | | | | |
| 460 | $6.748\times10^{21}$ | 4.48 | 0.15 | 0.0793 | 0.0115 | 0.09 | 0.5460 | 0.1663 |
| | | 4.01 | 0.14 | | | | | |
| | | 2.23 | 0.15 | | | | | |
| 480 | $5.389\times10^{21}$ | | | 0.0615 | 0.0136 | 0.08 | 0.4550 | 0.1650 |
| | | 2.75 | 0.13 | | | | | |
| | | 3.28 | 0.18 | | | | | |
| | | 1.16 | 0.09 | | | | | |
| 500 | $6.067\times10^{21}$ | 1.44 | 0.15 | 0.0256 | 0.009 | 0.03 | 0.3850 | 0.0898 |
| | | 1.27 | 0.11 | | | | | |
| | | 0.52 | 0.06 | | | | | |
| 520 | $6.686\times10^{21}$ | 0.42 | 0.11 | 0.0077 | 0.005 | 0.01 | 0.3360 | 0.0379 |
| | | 0.35 | 0.05 | | | | | |

**Table S4.** The average fluorescence lifetime of samples.

| Samples | $\tau_1$ (ns) | $B_1$ (%) | $\tau_2$ (ns) | $B_2$ (%) | Ave. $\tau$ (ns) |
|---|---|---|---|---|---|
| OIS | 1.0990 | 1610.21 | 11.1183 | 108.7143 | 1.7327 |
| OS | 1.1633 | 1619.26 | 10.8101 | 103.8460 | 1.7447 |
| IS | 1.2776 | 1489.63 | 6.9170 | 163.1824 | 1.8344 |